\documentclass[11pt]{article}

\usepackage[a4paper,margin=2.45cm]{geometry}
\usepackage{amsmath,amssymb,mathtools,bm,mathrsfs}
\usepackage{booktabs,array,longtable}
\usepackage{graphicx}
\usepackage{microtype}
\usepackage{enumitem}
\usepackage[numbers,sort&compress]{natbib}
\usepackage[colorlinks=true,allcolors=blue]{hyperref}
\usepackage[nameinlink,noabbrev]{cleveref}
\usepackage{xcolor}
\hypersetup{
  pdftitle={Compressible solved-volatility stochastic fluid thermodynamics: source-consistent energy, finite-correlation reservoirs, entropy admissibility and boundary conditions},
  pdfauthor={Hsieh-Chen Tsai},
  pdfsubject={Compressible stochastic fluid mechanics and thermodynamics},
  pdfkeywords={stochastic fluid mechanics, compressible flow, Ito-Hencky strain, entropy, finite-correlation reservoir}
}

\newcommand{\dd}{\mathrm d}
\newcommand{\D}{\mathrm D}
\newcommand{\tr}{\operatorname{tr}}
\newcommand{\sym}{\operatorname{sym}}

\newcommand{\dev}{\operatorname{dev}}
\newcommand{\Div}{\nabla\!\cdot}
\newcommand{\bI}{\bm I}
\newcommand{\bu}{\bm u}
\newcommand{\bSigmaAlpha}{\bm{\Sigma}_{\alpha}}
\newcommand{\bSigmaAlphaZero}{\bm{\Sigma}_{\alpha,0}}
\newcommand{\balpha}{\bm\alpha}
\newcommand{\bmom}{\bm m}
\newcommand{\bT}{\bm T}
\newcommand{\bq}{\bm q}
\newcommand{\bx}{\bm x}
\newcommand{\bn}{\bm n}
\newcommand{\bw}{\bm w}
\newcommand{\bz}{\bm z}
\newcommand{\bQ}{\bm Q}
\newcommand{\bD}{\bm D}
\newcommand{\bA}{\bm A}
\newcommand{\bL}{\bm L}
\newcommand{\bzero}{\bm 0}
\newcommand{\cE}{\mathcal E}
\newcommand{\cR}{\mathcal R}
\newcommand{\cU}{\mathcal U}
\newcommand{\cC}{\mathcal C}
\newcommand{\cF}{\mathcal F}

\newcommand{\cI}{\mathcal I}
\newcommand{\RR}{\mathbb R}
\newcommand{\E}{\mathbb E}

\title{\textbf{Compressible solved-volatility stochastic fluid thermodynamics:\\
source-consistent energy, finite-correlation reservoirs, entropy admissibility and boundary conditions}}
\author{Hsieh-Chen Tsai\\
Department of Mechanical Engineering, National Taiwan University\\
\texttt{hsiehchentsai@ntu.edu.tw}\\
ORCID: 0000-0002-4240-4332}
\date{30 July 2026}

\begin{document}
\maketitle

\begin{abstract}
A variable-density thermodynamic extension is developed for the solved-volatility stochastic-fluid formulation of arXiv:2607.25536. Source-inclusive stochastic transport separates mass, momentum and total-energy conservation into time-evolution partial differential equations and martingale compatibility constraints. Density and temperature are the primitive thermodynamic fields: mass conservation determines density, internal energy determines temperature, and the equation of state determines pressure evolution along stochastic particle paths. The resolved kinetic-energy identity is combined with a finite-correlation reservoir, Green--Kubo calibration, an equilibrium counterterm and adjoint resolved-unresolved exchange. A stochastic Gibbs identity and Gaussian relative entropy yield a conditional entropy-admissibility result for a Hencky-reservoir formulation. Equation-of-state pressure fluctuations are distinguished from mechanical stress impulses; regular finite-Mach fluctuations produce no independent white-noise bulk pressure impulse, while fast mechanical pressure is represented by a causal finite-correlation carrier. Conservative boundary conditions and a calorically perfect ideal-gas specialization are given. In the zero-volatility limit, the classical compressible Navier--Stokes--Fourier equations are recovered. A frozen descriptor analysis identifies a mixed hyperbolic--parabolic drift subsystem coupled to algebraic martingale constraints, with closure-dependent elliptic blocks and a singular low-Mach pressure limit. Canonical calculations verify the pressure carrier, acoustic dispersion, viscous-thermal energy balance and low-Mach scaling. Nonlinear well-posedness, shock admissibility and developed turbulence are not claimed.
\end{abstract}


\section{Introduction and relation to the incompressible theory}
\label{sec:intro}

Stochastic transport under location uncertainty provides conservative transport laws in which unresolved displacement enters through a spatial covariance, an It\^o--Stokes correction and stochastic source terms \citep{Memin2014,Resseguier2017,Tissot2026}.  Compressible mass, momentum and total-energy balances have already been derived in that setting \citep{Tissot2026}; the present paper does not claim priority for a stochastic compressible energy equation.  Its purpose is narrower: to combine source-consistent compressible transport with a solved displacement-volatility field, local It\^o--Hencky deformation, an explicit finite-correlation energy reservoir and a conditional entropy-admissibility result.

The route-independent local kinematics, one-channel source-consistent momentum formulation, index-one structure and finite-correlation boundary interpretation were developed in the companion incompressible formulation \citep{Tsai2026PaperI}, arXiv:2607.25536.  The present work extends that theory. It uses the same one-channel stochastic map but removes the constant-density and pathwise-isochoric restrictions.  The new questions are therefore the density--Jacobian coupling, thermodynamic pressure, total and internal energy, random--random power, entropy production and boundary energy fluxes.

The theoretical development follows the dependency chain of the field equations. The principal results are as follows.
\begin{enumerate}[leftmargin=2em]
\item The stochastic particle map and the It\^o--Hencky increment determine the logarithmic and current-volume rates, and hence the distinction between conservative transport and pressure--volume work.
\item Source-consistent mass, momentum and total-energy laws are written as time-evolution partial differential equations paired with martingale compatibility constraints. The resolved Eulerian fields have no independently assigned fixed-position martingale part.
\item For mass density $\rho$ and absolute temperature $T$, a thermodynamically compatible Helmholtz representation supplies the thermodynamic pressure $p(\rho,T)$ and specific internal energy $e(\rho,T)$, while the finite-Mach white-noise stress and deterministic Fourier heat flux provide the constitutive data used in the governing equations and their boundary fluxes.
\item It\^o differentiation of $K=|\bmom|^2/(2\rho)$, with $\bmom=\rho\bu$ and $\bu$ the resolved displacement drift, gives the exact resolved kinetic-energy identity. Subtraction from total energy then yields the physical internal--reservoir balance and the temperature evolution equation; equation-of-state differentiation gives the Eulerian pressure time derivative and the material pressure It\^o coefficients.
\item Random--random work is regularised at finite correlation time by a positive reservoir with Green--Kubo calibration, an equilibrium counterterm and adjoint resolved--unresolved exchange. A separate coloured pressure carrier is introduced only when fast mechanical pressure is physically retained.
\item A stochastic Gibbs identity and Gaussian relative-entropy calculation give a conditional entropy inequality for the Hencky--reservoir (H--R) formulation, together with the corresponding entropy and reservoir boundary conditions.
\item A complete calorically perfect ideal-gas specialization recovers the classical compressible Navier--Stokes--Fourier equations when the solved volatility and all associated stochastic amplitudes vanish.
\item Frozen-symbol and canonical-limit calculations distinguish finite-Mach hyperbolic--parabolic evolution, directional covariance rank, martingale compatibility, closure-dependent elliptic blocks and the singular low-Mach pressure constraint.
\end{enumerate}

The paper is organised in the same order. After the kinematics, the three conservation laws and their constitutive and boundary data are stated. The kinetic, internal and primitive thermodynamic equations are then derived. Finite-correlation regularisation and the optional pressure carrier precede the Gibbs and entropy analysis. The ideal-gas specialization, principal-symbol classification and canonical consistency calculations close the development.

The construction is one-channel and local in its It\^o--Hencky part. The one-channel covariance is directionally rank deficient in more than one spatial dimension, and the ellipticity of any algebraic solved-volatility block remains closure-dependent. Multichannel rough-path effects, shocks, global existence, nonlinear constitutive uniqueness and turbulence statistics remain outside the claims.

The conservative and thermodynamic balances obtained below do not by themselves provide universal constitutive closure. The H--R exchange must still be embedded in a finite-correlation reservoir. For the isotropic pressure channel, the analysis below closes the regular finite-Mach white-noise baseline by excluding an intrinsic bulk pressure impulse; an augmented coloured pressure carrier remains available when fast mechanical pressure is physically required. The formulation uses deterministic Fourier heat conduction and sets the bulk stochastic heat-flux impulse to zero. Any non-zero stochastic heat flux must be derived from an augmented thermal conservation law rather than postulated by multiplying Fourier's law by white noise.

\section*{Nomenclature and dimensional typing}
\addcontentsline{toc}{section}{Nomenclature and dimensional typing}
The Brownian increment has dimension $[\dd W_t]=\mathsf T^{1/2}$. In the dimension column, $M$, $L$, $\mathsf T$, $E$, $P$ and $\Theta_{\rm th}$ denote mass, length, time, energy, pressure and thermodynamic temperature, respectively; the field $T(\bx,t)$ denotes absolute temperature. Resolved Eulerian coefficient fields have finite variation at fixed position. Brownian dimensions enter the stochastic particle map, stress and source impulses, and material It\^o differentials.
{\small
\begin{longtable}{@{}p{0.18\textwidth}p{0.48\textwidth}p{0.23\textwidth}@{}}
\toprule
Symbol & Meaning & Dimension \\
\midrule
$\bm X_t,\ W_t$ & stochastic particle position and scalar Brownian motion & $L,\ \mathsf T^{1/2}$ \\
$\rho,\ T,\ p$ & mass density, absolute temperature and thermodynamic pressure & $ML^{-3},\ \Theta_{\rm th},\ P$ \\
$\bu$ & resolved It\^o displacement drift & $L\mathsf T^{-1}$ \\
$\balpha$ & solved displacement-volatility field & $L\mathsf T^{-1/2}$ \\
$\bSigmaAlpha=\balpha\otimes\balpha$ & displacement-covariance tensor per unit time & $L^2\mathsf T^{-1}$ \\
$\bI$ & identity tensor & dimensionless \\
$\bu_c,\bu^\star$ & conservative transport drift and current-volume work drift & $L\mathsf T^{-1}$ \\
$\bmom=\rho\bu$ & momentum density & $ML^{-2}\mathsf T^{-1}$ \\
$e,\ K,\cE$ & specific internal energy, resolved kinetic-energy density and total-energy density & $L^2\mathsf T^{-2},\ EL^{-3},\ EL^{-3}$ \\
$\theta_0,\Theta$ & logarithmic- and current-volume drift rates & $\mathsf T^{-1}$ \\
$\theta_1$ & volume martingale rate & $\mathsf T^{-1/2}$ \\
$\bm h_0,\bm h_1$ & It\^o--Hencky drift and martingale coefficients & $\mathsf T^{-1},\ \mathsf T^{-1/2}$ \\
$\bT_0,\bT_1$ & stress drift and stress-impulse coefficients & $P,\ P\mathsf T^{1/2}$ \\
$p_0^{\rm M},p_1^{\rm M}$ & material pressure drift and martingale coefficients along $\bm X_t$ & $P\mathsf T^{-1},\ P\mathsf T^{-1/2}$ \\
$\pi^\tau,\tau_p$ & finite-correlation mechanical pressure and its relaxation time & $P,\ \mathsf T$ \\
$K_s=\rho c_s^2$ & adiabatic bulk modulus & $P$ \\
$\bq_0,\bq_1$ & heat-flux drift and heat-flux-impulse coefficients & $\begin{gathered}EL^{-2}\mathsf T^{-1}\\ EL^{-2}\mathsf T^{-1/2}\end{gathered}$ \\
$\bz^\tau$ & finite-correlation unresolved velocity/reservoir state & $L\mathsf T^{-1}$ in the mechanical realisation \\
$\cR$ & renormalised unresolved-energy density & $EL^{-3}$ \\
$\mathcal H_0,\mathcal H_1$ & physical internal-energy drift-source and martingale-source amplitudes after reservoir subtraction & $EL^{-3}\mathsf T^{-1},\ EL^{-3}\mathsf T^{-1/2}$ \\
\bottomrule
\end{longtable}}
Bold $\bq$ is reserved for heat flux; the finite-correlation unresolved state is denoted by $\bz^\tau$. The white-noise map considered here has one scalar Brownian channel. Auxiliary finite-correlation wall or thermostat variables may have internal coordinates, but they do not introduce an additional retained displacement channel unless stated explicitly. The resolved coefficient fields may depend on adapted data, but no independent fixed-position $\dd W_t$ coefficient is assigned to $\rho$, $\bmom$, $T$ or $\cE$. Their Brownian channel appears as a compatibility equation generated by stochastic transport and impulse terms.

\section{Stochastic kinematics and volume change}
\label{sec:kinematics}

The particle map is
\begin{equation}
 \dd\bm X_t=\bu(\bm X_t,t)\,\dd t+\balpha(\bm X_t,t)\,\dd W_t,
 \label{eq:map}
\end{equation}
where $W_t$ is a scalar Brownian motion, $\bu$ is the resolved It\^o displacement drift and $\balpha$ is the solved displacement-volatility field.  Set
\begin{equation}
 \bSigmaAlpha=\balpha\otimes\balpha,
 \qquad
 \bL=\nabla\bu,
 \qquad
 \bA=\nabla\balpha,
 \qquad
 \bD_u=\sym\bL,
 \qquad
 \bD_\alpha=\sym\bA,
\end{equation}
where $\sym\bm B=(\bm B+\bm B^{\mathsf T})/2$. Let $\bm h$ denote the local spatial Hencky strain. The It\^o--Hencky increment derived in \citet{Tsai2026PaperI}, in the logarithmic-strain tradition of \citet{Xiao1997} and \citet{Norris2008}, is
\begin{equation}
 \dd\bm h=\bm h_0\,\dd t+\bm h_1\,\dd W_t,
 \qquad
 \bm h_1=\bD_\alpha,
 \qquad
 \bm h_0=\bD_u+\frac12\bA\bA^{\mathsf T}-\bD_\alpha^2.
 \label{eq:hencky}
\end{equation}
Let $J_t=\det(\partial\bm X_t/\partial\bm X_0)$ be the local volume Jacobian of the stochastic flow. Taking the trace gives
\begin{equation}
 \dd\log J=\theta_0\,\dd t+\theta_1\,\dd W_t,
 \qquad
 \theta_1=\Div\balpha,
 \qquad
 \theta_0=\Div\bu-\frac12\tr(\bA^2).
 \label{eq:logJ}
\end{equation}
It\^o's formula distinguishes the drift of $\log J$ from the actual drift of $J$:
\begin{equation}
 \frac{\dd J}{J}=\Theta\,\dd t+\theta_1\,\dd W_t,
 \qquad
 \Theta=\theta_0+\frac12\theta_1^2.
 \label{eq:Jratio}
\end{equation}
With
\begin{equation}
 \bu_c=\bu-\frac12\Div\bSigmaAlpha,
 \qquad
 \bu^\star=\bu_c+\balpha(\Div\balpha),
 \label{eq:drifts}
\end{equation}
one obtains, for one channel,
\begin{equation}
 \Div\bu^\star=\Theta.
 \label{eq:ustardiv}
\end{equation}
The conservative transport laws use $\bu_c$; primitive pressure-volume work uses $\bu^\star$.  They coincide only under additional restrictions.

For a material parcel of fixed mass, let $\rho$ be mass density and $v=1/\rho$ the specific volume. Then
\begin{equation}
 \dd v=v\Theta\,\dd t+v\theta_1\,\dd W_t,
 \label{eq:vSDE}
\end{equation}
which is equivalent to $\rho(\bm X_t,t)J_t=\text{constant}$.

\section{Source-consistent governing equations}
\label{sec:balances}

Let $\Omega\subset\RR^d$ be the flow domain in spatial dimension $d$, with position $\bx\in\Omega$. The resolved Eulerian coefficient fields $\rho(\bx,t)$, $\bu(\bx,t)$, $T(\bx,t)$, $\balpha(\bx,t)$ and $\cR(\bx,t)$ are taken to have finite variation at each fixed spatial point. Thus, for any such scalar or vector field $y$,
\begin{equation}
 \dd_t y(\bx,t)=\partial_t y(\bx,t)\,\dd t.
 \label{eq:finitevariationfield}
\end{equation}
There is no independent fixed-position martingale coefficient of $y$. The $\dd W_t$ channel of each conservation law is generated by stochastic transport and impulse terms and becomes a spatial compatibility constraint after coefficient comparison. Along the stochastic particle path \cref{eq:map}, however, It\^o's formula gives, componentwise,
\begin{equation}
\begin{aligned}
 \dd y(\bm X_t,t)={}&
 \left[\partial_t y+\bu\cdot\nabla y+
 \frac12\bSigmaAlpha:\nabla^2y\right]_{\bm X_t}\dd t
 +\left[\balpha\cdot\nabla y\right]_{\bm X_t}\dd W_t.
\end{aligned}
\label{eq:materialItoResolved}
\end{equation}
Equation \eqref{eq:materialItoResolved} is where the material martingale coefficient appears; it is not an independently assigned Eulerian state variable.

The primary conserved densities are mass density $\rho$, momentum density $\bmom=\rho\bu$ and total-energy density $\cE$. Temperature is obtained from the physical internal-energy balance in \cref{sec:temperature}, and thermodynamic pressure is obtained from the equation of state after $\rho$ and $T$ have been determined.

\subsection{Mass conservation}

Comparing the finite-variation and Brownian channels gives
\begin{subequations}
\label{eq:mass}
\begin{align}
 \partial_t\rho+\Div(\rho\bu_c)
 &=\frac12\Div(\bSigmaAlpha\nabla\rho),
 \label{eq:massdrift}\\
 \Div(\rho\balpha)&=0.
 \label{eq:massmart}
\end{align}
\end{subequations}
The first equation is the mass-evolution partial differential equation. The second is the martingale mass-compatibility constraint. In particular,
\begin{equation}
 \balpha\cdot\nabla\rho=-\rho\theta_1.
 \label{eq:rhoAlphaConstraint}
\end{equation}
Substitution of \cref{eq:mass} into the full material It\^o differential \cref{eq:materialItoResolved} gives
\begin{subequations}
\label{eq:materialrho}
\begin{align}
 \dd\rho(\bm X_t,t)
 &=\rho\big(\theta_1^2-\Theta\big)\,\dd t
   -\rho\theta_1\,\dd W_t,
 \label{eq:materialrho-direct}\\
 \dd\log\rho(\bm X_t,t)
 &=-\theta_0\,\dd t-\theta_1\,\dd W_t.
 \label{eq:materialrho-log}
\end{align}
\end{subequations}
Together with \cref{eq:logJ}, this yields $\dd\log(\rho J)=0$ and hence parcel-mass conservation.

The pathwise-isochoric limit associated with the companion incompressible formulation follows directly. If $\rho$ is constant, \cref{eq:massmart} gives $\Div\balpha=0$, while \cref{eq:massdrift} gives $\Div\bu_c=0$. Using $\bu_c=\bu-\tfrac12\Div(\balpha\otimes\balpha)$ then yields
\begin{equation}
 \Div\bu=\frac12\tr\!\left[(\nabla\balpha)^2\right],
 \label{eq:paperIisoclimit}
\end{equation}
which is the compressible-to-isochoric correspondence with equation (4) of arXiv:2607.25536.

\subsection{Momentum balance}

Let $\bm b_0$ and $\bm b_1$ be the body-force drift and body-force-impulse amplitudes per unit mass. The mechanical source coefficients are
\begin{equation}
 \bQ_0=\Div\bT_0+\rho\bm b_0,
 \qquad
 \bQ_1=\Div\bT_1+\rho\bm b_1.
 \label{eq:Qdef}
\end{equation}
For the energy decomposition below, write
\begin{equation}
 \bm b_0=-\nabla\Phi+\bm b_0^{nc},
 \qquad
 \bm b_1=\bm b_1^{nc},
 \label{eq:bodyforcesplit}
\end{equation}
where $\Phi$ is a deterministic conservative potential per unit mass.

Momentum conservation separates into
\begin{subequations}
\label{eq:momentum}
\begin{align}
 \partial_t\bmom
 +\Div(\bmom\otimes\bu_c)
 +\Div(\bQ_1\otimes\balpha)
 &=\frac12\Div(\bSigmaAlpha\nabla\bmom)+\bQ_0,
 \label{eq:momentumdrift}\\
 \Div(\bmom\otimes\balpha)&=\bQ_1.
 \label{eq:momentummart}
\end{align}
\end{subequations}
The first line is the momentum-evolution partial differential equation; the second is the martingale momentum-compatibility equation. In the constant-density pathwise-isochoric limit, $\Div\balpha=0$ and \cref{eq:momentummart} reduces to
\begin{equation}
 \bQ_1=\rho(\balpha\cdot\nabla)\bu,
 \label{eq:paperImomentumlimit}
\end{equation}
the martingale momentum relation used in the companion incompressible formulation.

\subsection{Total-energy balance}

Define the resolved kinetic-energy density $K$, specific internal energy $e$, and total-energy density $\cE$ by
\begin{equation}
 K=\frac{|\bmom|^2}{2\rho},
 \qquad
 \cE=\rho e(\rho,T)+\cR+K+\rho\Phi,
 \label{eq:Edef}
\end{equation}
where $\cR$ is the finite-correlation/renormalised unresolved-energy density. The white-noise displacement $\balpha\dd W_t$ is not assigned the ordinary energy $\rho|\balpha|^2/2$.

Let $\bq_0,\bq_1$ denote the heat-flux drift and heat-flux-impulse amplitudes, and let $r_0\dd t+r_1\dd W_t$ denote the local volumetric energy supply. With the non-conservative body-force channels of \cref{eq:bodyforcesplit}, define
\begin{equation}
\begin{aligned}
 R_0^E={}&\Div(\bT_0\bu^\star)-\Div\bq_0
 +\rho\bm b_0^{nc}\cdot\bu^\star+r_0,\\
 R_1^E={}&\Div(\bT_0\balpha+\bT_1\bu^\star-\bq_1)
 +\rho\bm b_0^{nc}\cdot\balpha
 +\rho\bm b_1^{nc}\cdot\bu^\star+r_1.
\end{aligned}
\label{eq:Esource}
\end{equation}
These are the non-transport energy-source amplitudes, not separate balances. No naked random--random product $\bT_1:\nabla\balpha\,\dd t$ is added, because that product belongs to the finite-correlation reservoir.

Total-energy conservation separates into
\begin{subequations}
\label{eq:Econservative}
\begin{align}
 \partial_t\cE+\Div(\cE\bu_c)+\Div(\balpha R_1^E)
 &=\frac12\Div(\bSigmaAlpha\nabla\cE)+R_0^E,
 \label{eq:Edrift}\\
 \Div(\cE\balpha)&=R_1^E.
 \label{eq:Emart}
\end{align}
\end{subequations}
The source--transport covariance $\Div(\balpha R_1^E)$ belongs to the drift equation, while the second line is the martingale energy-compatibility constraint.

\begin{table}[ht]
\centering
\caption{Primary finite-variation evolution equations and martingale compatibility constraints.}
\label{tab:primarychannels}
\small
\renewcommand{\arraystretch}{1.55}
\begin{tabular}{@{}>{\centering\arraybackslash}p{0.10\textwidth}>{\raggedright\arraybackslash}p{0.45\textwidth}>{\raggedright\arraybackslash}p{0.35\textwidth}@{}}
\toprule
Resolved field & Drift evolution equation & Martingale compatibility constraint \\
\midrule
$\rho$
& $\displaystyle \partial_t\rho+\Div(\rho\bu_c)=\frac12\Div(\bSigmaAlpha\nabla\rho)$
& $\displaystyle \Div(\rho\balpha)=0$ \\
$\bmom=\rho\bu$
& $\displaystyle \partial_t\bmom+\Div(\bmom\otimes\bu_c)+\Div(\bQ_1\otimes\balpha)=\frac12\Div(\bSigmaAlpha\nabla\bmom)+\bQ_0$
& $\displaystyle \Div(\bmom\otimes\balpha)=\bQ_1$ \\
$\cE$
& $\displaystyle \partial_t\cE+\Div(\cE\bu_c)+\Div(\balpha R_1^E)=\frac12\Div(\bSigmaAlpha\nabla\cE)+R_0^E$
& $\displaystyle \Div(\cE\balpha)=R_1^E$ \\
\bottomrule
\end{tabular}
\end{table}

For compact derived balances later in the paper, define the source-consistent scalar operator
\begin{equation}
\begin{aligned}
 \cC_s[q;R_1]={}&\partial_t q\,\dd t
 +\Div\left[q(\bu_c\dd t+\balpha\dd W_t)\right]
 +\Div(\balpha R_1)\dd t
 -\frac12\Div(\bSigmaAlpha\nabla q)\dd t,
\end{aligned}
\label{eq:Cs}
\end{equation}
and, for a vector extensive density $\bm q$ with martingale source coefficient $\bm R_1$,
\begin{equation}
\begin{aligned}
 \cC_v[\bm q;\bm R_1]={}&\partial_t\bm q\,\dd t
 +\Div\left[\bm q\otimes(\bu_c\dd t+\balpha\dd W_t)\right]
 +\Div(\bm R_1\otimes\balpha)\dd t
 -\frac12\Div(\bSigmaAlpha\nabla\bm q)\dd t.
\end{aligned}
\label{eq:Cv}
\end{equation}
The source--transport quadratic-covariation terms in these operators are already displayed explicitly in \cref{tab:primarychannels}. Removing them changes both the drift contribution and the conservative boundary flux.

\section{Thermodynamic closure and white-noise constitutive laws}
\label{sec:stress}

\subsection{Primitive thermodynamic fields and Helmholtz representation}

The primitive thermodynamic fields are $\rho(\bx,t)$ and $T(\bx,t)$. A thermodynamically consistent single-component closure may be generated from a specific Helmholtz free energy $\psi=\psi(\rho,T)$; the associated $s$, $p$ and $e$ below are specific entropy, thermodynamic pressure and specific internal energy.
\begin{equation}
 s=-\psi_T,
 \qquad
 p=\rho^2\psi_\rho,
 \qquad
 e=\psi+Ts,
 \label{eq:helmholtz}
\end{equation}
with $T>0$ and $e_T=c_v>0$. Equivalent stable equations of state may be used, but $p(\rho,T)$ and $e(\rho,T)$ must satisfy the Gibbs/Maxwell compatibility encoded by \cref{eq:helmholtz}. Mass conservation determines $\dd_t\rho$; the physical internal-energy balance determines $\dd_tT$ in \cref{sec:temperature}; and the pressure-state differential follows from the equation of state rather than from an independent pressure evolution law.

\subsection{Finite-Mach It\^o--Hencky stress}

Let $\mu\ge0$ and $\zeta\ge0$ be the shear and bulk viscosities and let $\bI$ be the identity tensor. The finite-Mach white-noise stress used in the momentum equation is
\begin{equation}
\begin{aligned}
 \bT_0&=-p\bI+2\mu\dev\bm h_0+\zeta\Theta\bI,\\
 \bT_1&=2\mu\dev\bm h_1+\zeta\theta_1\bI.
\end{aligned}
\label{eq:stress}
\end{equation}
Here $p=p(\rho,T)$ is the ordinary thermodynamic pressure obtained from the thermodynamic potential. Mixtures require additional species balances and are outside the present closure. The regular finite-Mach white-noise baseline contains no intrinsic isotropic pressure impulse in $\bT_1$; the martingale bulk stress shown here is the viscous impulse $\zeta\theta_1\bI$. A retained finite-correlation pressure carrier is derived after the general reservoir regularisation. The drift bulk rate is the current-volume rate $\Theta=\Div\bu^\star$, not the drift $\theta_0$ of $\log J$.

The drift stress in \cref{eq:stress} is the mechanical constitutive law used in momentum. Its direct power against the current-volume rate is not pointwise sign-definite. The exact Hencky--reservoir power partition is derived only after the resolved kinetic and internal--reservoir balances have been obtained in \cref{sec:energy}; its entropy interpretation is then established in \cref{sec:entropy}.

\subsection{Heat flux and baseline stochastic channels}

Let $\kappa\ge0$ be the thermal conductivity. The finite-Mach white-noise formulation studied here uses
\begin{equation}
 \bq_0=-\kappa\nabla T,
 \qquad
 \bq_1=\bzero.
 \label{eq:Fourier}
\end{equation}
Thus the bulk conductive heat flux is deterministic, while stochasticity still enters energy through transport, stress work and resolved--reservoir exchange. A non-zero heat-flux martingale is not obtained by multiplying Fourier's law by white noise; it requires an augmented thermal conservation law with its own carrier energy, counterterm, entropy and boundary fluxes.
\section{Boundary conditions for the governing equations}
\label{sec:bc}

Boundary data must be imposed on the complete conservative fluxes associated with the drift and martingale equations, rather than on selected stress or transport terms. The formulas below are stated first on a fixed boundary $\Gamma$ with outward normal $\bn$.

\subsection{Mass, momentum and total-energy fluxes}

The outward mass fluxes corresponding to \cref{eq:mass} are
\begin{equation}
 \bm\cF_\rho^0=\rho\bu_c-\frac12\bSigmaAlpha\nabla\rho,
 \qquad
 \bm\cF_\rho^1=\rho\balpha.
 \label{eq:massflux}
\end{equation}
The outward momentum fluxes corresponding to \cref{eq:momentum} are
\begin{equation}
 \bm\cF_m^0=\bmom\otimes\bu_c+\bQ_1\otimes\balpha-\frac12\bSigmaAlpha\nabla\bmom-\bT_0,
 \qquad
 \bm\cF_m^1=\bmom\otimes\balpha-\bT_1.
 \label{eq:momflux}
\end{equation}
The outward total-energy fluxes corresponding to \cref{eq:Econservative} are
\begin{equation}
\begin{aligned}
 \bm\cF_E^0={}&\cE\bu_c+\balpha R_1^E-\frac12\bSigmaAlpha\nabla\cE
 -\bT_0\bu^\star+\bq_0,\\
 \bm\cF_E^1={}&\cE\balpha-\bT_0\balpha-\bT_1\bu^\star+\bq_1.
\end{aligned}
\label{eq:Eflux}
\end{equation}
Thus a boundary condition for each conservation law has both a drift flux and, when stochastic transport or impulse is present, a martingale flux. The stress and deterministic Fourier heat-flux coefficients in these formulas are specified in \cref{sec:stress}.

For a general scalar extensive balance with $R_0=\Div\bm F_0+s_0$ and $R_1=\Div\bm F_1+s_1$, the same construction gives
\begin{equation}
 \bm\cF_q^0=q\bu_c+\balpha R_1-\frac12\bSigmaAlpha\nabla q-\bm F_0,
 \qquad
 \bm\cF_q^1=q\balpha-\bm F_1.
 \label{eq:scalarflux}
\end{equation}

\subsection{Deterministically moving boundaries}

For a boundary moving with deterministic velocity $\bw_\Gamma$, the drift fluxes relative to the boundary are
\begin{equation}
 \bm\cF_{\rho,\Gamma}^0=\bm\cF_\rho^0-\rho\bw_\Gamma,
 \qquad
 \bm\cF_{m,\Gamma}^0=\bm\cF_m^0-\bmom\otimes\bw_\Gamma,
 \qquad
 \bm\cF_{E,\Gamma}^0=\bm\cF_E^0-\cE\bw_\Gamma.
 \label{eq:movingflux}
\end{equation}
The martingale fluxes are unchanged when the boundary motion is deterministic. Omitting the swept-volume terms in \cref{eq:movingflux} violates the Reynolds transport balance.

\subsection{Impermeable material wall}

Let $\varphi(\bx,t)=0$ be a signed-distance representation of a deterministic wall moving with velocity $\bw_\Gamma$. Pathwise invariance of the stochastic particle map gives
\begin{equation}
 \balpha\cdot\bn=0,
 \qquad
 (\bu-\bw_\Gamma)\cdot\bn+\frac12\bSigmaAlpha:\nabla\bn=0.
\end{equation}
Using $\bn\cdot\Div\bSigmaAlpha=-\bSigmaAlpha:\nabla\bn$ for a tangential covariance field, this is equivalent to
\begin{equation}
 \balpha\cdot\bn=0,
 \qquad
 (\bu_c-\bw_\Gamma)\cdot\bn=0.
 \label{eq:impermeable}
\end{equation}
Because $\bSigmaAlpha\bn=\balpha(\balpha\cdot\bn)=\bzero$, the normal covariance-diffusion mass flux also vanishes. A no-slip drift wall additionally imposes
\begin{equation}
 (\bI-\bn\otimes\bn)(\bu-\bw_\Gamma)=\bzero.
\end{equation}
An overbar denotes prescribed boundary data. The tangential volatility is a separate physical choice:
\begin{enumerate}[label=(\alph*),leftmargin=2em]
\item \emph{fully clamped stochastic wall}: $\balpha=\bzero$;
\item \emph{active receptivity}: prescribe $(\bI-\bn\otimes\bn)\balpha=\bar{\balpha}_{\rm tan}$ and include the required boundary work;
\item \emph{passive tangential wall}: retain $\balpha\cdot\bn=0$ and determine the tangential trace from an impedance/fluctuation--dissipation law;
\item \emph{martingale traction}: prescribe $(\bI-\bn\otimes\bn)\bT_1\bn$ rather than the trace of $\balpha$.
\end{enumerate}
Wall tangency bounds the covariance range but does not determine the number of possible stochastic modes \citep{Tsai2026PaperI}.

\subsection{Mechanical traction and open boundaries}

On a traction boundary, the mechanically natural data are the complete momentum fluxes
\begin{equation}
 \bm\cF_m^0\bn=\bar{\bm f}_m^0,
 \qquad
 \bm\cF_m^1\bn=\bar{\bm f}_m^1.
 \label{eq:totaltraction}
\end{equation}
On a stationary impermeable wall the transport terms simplify and these conditions reduce to Cauchy traction and traction-impulse data. On an open boundary they do not: advective, covariance-diffusive and source-covariation contributions remain. For a subsonic deterministic inflow/outflow, the number of prescribed drift variables follows the incoming characteristics of the compressible Navier--Stokes--Fourier part. Stochastic inflow data $\balpha\cdot\bn$, $\bT_1\bn$ and thermodynamic state fluctuations must additionally satisfy the mass, energy and pressure-channel compatibility conditions.

\subsection{Thermal boundary conditions}

Let $\Gamma_T$ and $\Gamma_q$ denote boundary subsets on which temperature or conductive heat flux is prescribed. The conductive thermal data may be specified as
\begin{equation}
 T=\bar T\quad\text{on }\Gamma_T,
 \qquad\text{or}\qquad
 \bq_0\cdot\bn=\bar q_0\quad\text{on }\Gamma_q.
 \label{eq:thermalbc}
\end{equation}
An adiabatic boundary has $\bq_0\cdot\bn=0$. The present bulk formulation sets $\bq_1=\bzero$. A prescribed stochastic boundary heat flux is an external martingale energy flux and must be introduced together with its source--transport covariation and its entropy contribution; it is not obtained by multiplying Fourier's law by white noise.

If the finite-correlation pressure carrier \cref{eq:pressurecarrier} is retained, it is an additional incoming relaxation variable at an open boundary. One may prescribe the incoming value of $\pi^\tau$ or a dissipative pressure impedance, but not both. Its mechanical energy flux is $\pi^\tau\bz^\tau\cdot\bn$. At an impermeable closed wall with $\bz^\tau\cdot\bn=0$, this flux vanishes and the local relaxation energy remains part of the reservoir balance.

\subsection{Closed mass and energy boundary conditions}

For a fixed closed boundary, a sufficient conservative statement of closed mass and total energy is
\begin{equation}
 \bm\cF_\rho^0\cdot\bn=0,
 \qquad
 \bm\cF_\rho^1\cdot\bn=0,
 \qquad
 \bm\cF_E^0\cdot\bn=0,
 \qquad
 \bm\cF_E^1\cdot\bn=0.
 \label{eq:closedmassenergy}
\end{equation}
On a deterministic moving boundary, the relative drift fluxes in \cref{eq:movingflux} replace the fixed-boundary drift fluxes. The additional entropy and finite-correlation reservoir conditions required for a closed adiabatic system are given after the entropy balance in \cref{sec:entropybc}.

\section{Resolved kinetic energy}
\label{sec:kinetic}

Let
\begin{equation}
 K=\frac{|\bmom|^2}{2\rho}.
\end{equation}
Because $\rho$ and $\bmom$ have finite variation at fixed position, their local chain rule is
\begin{equation}
 \partial_tK=\bu\cdot\partial_t\bmom
 -\frac12|\bu|^2\partial_t\rho.
 \label{eq:Klocaltime}
\end{equation}
There is no intrinsic fixed-position It\^o Hessian term in \cref{eq:Klocaltime}. The quadratic-variation contribution in the conservative kinetic-energy balance instead arises when the martingale momentum source, stochastic transport and source--transport covariance are combined.

Using \cref{eq:mass,eq:momentum} and the source-consistent operator \cref{eq:Cs}, one obtains
\begin{equation}
 \cC_s[K;\bu\cdot\bQ_1]
 =\left(\bu\cdot\bQ_0+\frac{|\bQ_1|^2}{2\rho}\right)\dd t
 +(\bu\cdot\bQ_1)\dd W_t.
 \label{eq:Kformulation}
\end{equation}
The martingale compatibility equations imply
\begin{equation}
 \bQ_1=\Div(\bmom\otimes\balpha)
 =\rho(\balpha\cdot\nabla)\bu,
 \label{eq:Q1onshell}
\end{equation}
where the second equality uses $\Div(\rho\balpha)=0$. Thus the fixed-position momentum martingale vanishes on the constraint manifold even though the material/source representation retains the quadratic-variation exchange $|\bQ_1|^2/(2\rho)$.

To see the representation equivalence, let $\bm G=(\balpha\cdot\nabla)\bu$ and, more generally, let $\bm\chi_1$ denote the non-transport momentum impulse after any mass-source correction. The local Eulerian Hessian term, source--transport covariance and stochastic-diffusion chain rule combine as
\begin{equation}
 \frac{|\bm\chi_1-\rho\bm G|^2}{2\rho}
 +\bm\chi_1\cdot\bm G
 -\frac\rho2|\bm G|^2
 =\frac{|\bm\chi_1|^2}{2\rho}.
 \label{eq:centralcancel}
\end{equation}
On the present no-mass-source constraint manifold, $\bm\chi_1=\bQ_1=\rho\bm G$. The positive term in \cref{eq:Kformulation} is therefore a resolved kinetic-energy quadratic-variation input or inter-sector exchange, not a thermodynamic dissipation and not an independent Eulerian noise amplitude.

\section{Internal--reservoir balance}
\label{sec:energy}

The total-energy law has already been stated in \cref{eq:Econservative}. The resolved kinetic-energy identity of \cref{sec:kinetic} permits an exact separation of the physical internal energy and the finite-correlation reservoir from the total balance.

The potential-energy balance follows from mass conservation.  With
\begin{equation}
 R_1^P=\rho\balpha\cdot\nabla\Phi,
 \qquad
 R_0^P=\rho\left(\bu\cdot\nabla\Phi+\frac12\bSigmaAlpha:\nabla^2\Phi\right),
\end{equation}
one has $\cC_s[\rho\Phi;R_1^P]=R_0^P\dd t+R_1^P\dd W_t$.

Subtracting the kinetic and potential equations from \cref{eq:Econservative} gives the exact combined internal--reservoir equation
\begin{equation}
 \cC_s[\cU;R_1^U]=R_0^U\dd t+R_1^U\dd W_t,\qquad \cU=\rho e+\cR.
 \label{eq:Uintro}
\end{equation}
If $\bm d=\bu^\star-\bu$ and $\cI_K=|\bQ_1|^2/(2\rho)$, then
\begin{equation}
\begin{aligned}
 R_0^U={}&\bT_0:\nabla\bu^\star+\bm d\cdot\bQ_0
 +\rho\bm d\cdot\nabla\Phi-\frac12\rho\bSigmaAlpha:\nabla^2\Phi\\
 &-\Div\bq_0+r_0-\cI_K,
\end{aligned}
\label{eq:Udrift}
\end{equation}
\begin{equation}
\begin{aligned}
 R_1^U={}&\bT_0:\nabla\balpha+\bT_1:\nabla\bu^\star
 +\balpha\cdot\bQ_0+\bm d\cdot\bQ_1
 -\Div\bq_1+r_1.
\end{aligned}
\label{eq:Umart}
\end{equation}
The kinetic quadratic-variation input $+\cI_K$ appears as $-\cI_K$ in the internal--reservoir sector, so total energy is not created twice.

\subsection{Hencky--reservoir mechanical power partition}

Define
\begin{equation}
 \overline{\bD}^{\star}=\dev\sym\nabla\bu^\star,
 \qquad
 \overline{\bm h}_0=\dev\bm h_0,
 \qquad
 \bm\Delta_H=\overline{\bD}^{\star}-\overline{\bm h}_0.
\end{equation}
The mechanical drift power then separates exactly as
\begin{equation}
 \bT_0:\nabla\bu^\star
 =-p\Theta
 +\underbrace{2\mu|\overline{\bm h}_0|^2+\zeta\Theta^2}_{\Phi_H\ge0}
 +\underbrace{2\mu\overline{\bm h}_0:\bm\Delta_H}_{\mathscr X_H}.
 \label{eq:HRsplit}
\end{equation}
The Hencky--reservoir formulation assigns $\Phi_H$ to thermodynamic heating and $\mathscr X_H$ to reversible finite-correlation reservoir work. This partition changes neither momentum nor total energy. Its reservoir embedding and entropy admissibility are stated in \cref{eq:HRembedding,eq:entropyprod}.

\subsection{Temperature equation derived from physical internal energy}
\label{sec:temperature}

Let $\varepsilon=\rho e(\rho,T)$ denote the physical internal-energy density. A local reservoir realisation is written in source-consistent form as
\begin{subequations}
\label{eq:reservoirlocalchannels}
\begin{align}
 \partial_t\cR+\Div(\cR\bu_c)+\Div(\balpha S_1^{\cR})
 &=\frac12\Div(\bSigmaAlpha\nabla\cR)+S_0^{\cR},
 \label{eq:reservoirlocaldrift}\\
 \Div(\cR\balpha)&=S_1^{\cR},
 \label{eq:reservoirlocalmart}
\end{align}
\end{subequations}
where $S_0^{\cR}$ and $S_1^{\cR}$ are the resolved drift-source and martingale-source amplitudes supplied by the finite-correlation closure. Define the physical internal-energy source pair
\begin{equation}
 \mathcal H_0=R_0^U-S_0^{\cR},
 \qquad
 \mathcal H_1=R_1^U-S_1^{\cR}.
 \label{eq:H01def}
\end{equation}
Subtracting \cref{eq:reservoirlocalchannels} from the combined internal--reservoir balance gives
\begin{subequations}
\label{eq:physicalinternalchannels}
\begin{align}
 \partial_t\varepsilon+\Div(\varepsilon\bu_c)
 +\Div(\balpha\mathcal H_1)
 &=\frac12\Div(\bSigmaAlpha\nabla\varepsilon)+\mathcal H_0,
 \label{eq:physicalinternaldrift}\\
 \Div(\varepsilon\balpha)&=\mathcal H_1.
 \label{eq:physicalinternalmart}
\end{align}
\end{subequations}
The first line determines the time evolution of physical internal energy; the second is its martingale compatibility condition after the reservoir exchange has been removed.

Because the resolved fields have finite variation at fixed position,
\begin{equation}
 \partial_t\varepsilon
 =\varepsilon_\rho\,\partial_t\rho
 +\varepsilon_T\,\partial_tT,
 \qquad \varepsilon_T>0.
 \label{eq:epsilonTimeChain}
\end{equation}
Using \cref{eq:massdrift,eq:physicalinternaldrift}, the temperature evolution equation is therefore
\begin{equation}
\begin{aligned}
 \partial_tT=\frac{1}{\varepsilon_T}\Bigg\{&
 -\Div(\varepsilon\bu_c)-\Div(\balpha\mathcal H_1)
 +\frac12\Div(\bSigmaAlpha\nabla\varepsilon)+\mathcal H_0\\
 &-\varepsilon_\rho\left[-\Div(\rho\bu_c)
 +\frac12\Div(\bSigmaAlpha\nabla\rho)\right]\Bigg\}.
\end{aligned}
\label{eq:Tdriftgeneral}
\end{equation}
The martingale compatibility equations \cref{eq:massmart,eq:physicalinternalmart} give the corresponding spatial relation
\begin{equation}
 \varepsilon_T\,\balpha\cdot\nabla T
 =\mathcal H_1+\big(\rho\varepsilon_\rho-\varepsilon\big)\theta_1.
 \label{eq:Tmartgeneral}
\end{equation}
Thus $T(\bx,t)$ is not assigned an independent stochastic conservation law. Its time derivative follows from physical internal energy, while its material martingale coefficient is the spatial quantity $\balpha\cdot\nabla T$ fixed by \cref{eq:Tmartgeneral}. Along the particle path,
\begin{equation}
 \dd T(\bm X_t,t)=
 \left[\partial_tT+\bu\cdot\nabla T+\frac12\bSigmaAlpha:\nabla^2T\right]\dd t
 +(\balpha\cdot\nabla T)\dd W_t.
 \label{eq:materialTIto}
\end{equation}

The mechanical part of the drift stress power has the exact split
\begin{equation}
 \bT_0:\nabla\bu^\star=-p\Theta+\Phi_H+\mathscr X_H.
 \label{eq:mechanicalsplit}
\end{equation}
The H--R closure is the explicit exchange matching
\begin{equation}
 R_{0,\mathrm{mech}}^{e}=-p\Theta+\Phi_H,
 \qquad
 R_{0,\mathrm{mech}}^{\cR}=\mathscr X_H,
 \qquad
 R_{0,\mathrm{mech}}^{e}+R_{0,\mathrm{mech}}^{\cR}=\bT_0:\nabla\bu^\star.
 \label{eq:HRmatching}
\end{equation}
Thus $\mathscr X_H$ is not deleted and is not counted as heat. It is transferred to the reservoir with the opposite sign in the resolved mechanical energy balance. The kinetic quadratic variation, pressure half-bracket and source-alignment terms are treated analogously as reversible inter-sector exchanges unless an independent constitutive law identifies a genuinely dissipative part.

\subsection{Pressure-state evolution and material It\^o coefficients from the equation of state}

At fixed position, the equation of state $p=p(\rho,T)$ gives the ordinary time-chain rule
\begin{equation}
 \partial_t p=p_\rho\,\partial_t\rho+p_T\,\partial_tT,
 \label{eq:pEulerianTime}
\end{equation}
where $\partial_t\rho$ and $\partial_tT$ are supplied by \cref{eq:massdrift,eq:Tdriftgeneral}. The pressure martingale is not an independently assigned Eulerian coefficient. It appears in the material It\^o differential
\begin{equation}
 \dd p(\bm X_t,t)=p_0^{\rm M}\,\dd t+p_1^{\rm M}\,\dd W_t,
 \label{eq:pMaterialSDE}
\end{equation}
with
\begin{subequations}
\label{eq:pMaterialCoefficients}
\begin{align}
 p_0^{\rm M}
 &=\partial_t p+\bu\cdot\nabla p+\frac12\bSigmaAlpha:\nabla^2p,
 \label{eq:pMaterialDrift}\\
 p_1^{\rm M}
 &=\balpha\cdot\nabla p
 =p_\rho\,\balpha\cdot\nabla\rho
  +p_T\,\balpha\cdot\nabla T.
 \label{eq:pMaterialMart}
\end{align}
\end{subequations}
Using \cref{eq:rhoAlphaConstraint,eq:Tmartgeneral}, the EOS-derived material pressure martingale is
\begin{equation}
 p_1^{\rm M}
 =-\rho p_\rho\theta_1
 +\frac{p_T}{\varepsilon_T}
 \left[\mathcal H_1+(\rho\varepsilon_\rho-\varepsilon)\theta_1\right].
 \label{eq:pMaterialMartClosed}
\end{equation}
The It\^o curvature in \cref{eq:pMaterialDrift} is also determined by the equation of state. Since $\bSigmaAlpha=\balpha\otimes\balpha$,
\begin{equation}
\begin{aligned}
 \bSigmaAlpha:\nabla^2p={}&p_\rho\,\bSigmaAlpha:\nabla^2\rho
 +p_T\,\bSigmaAlpha:\nabla^2T
 +p_{\rho\rho}(\balpha\cdot\nabla\rho)^2\\
 &+2p_{\rho T}(\balpha\cdot\nabla\rho)(\balpha\cdot\nabla T)
 +p_{TT}(\balpha\cdot\nabla T)^2.
\end{aligned}
\label{eq:pEOSCurvature}
\end{equation}
Hence both material pressure channels follow from mass conservation, physical internal-energy conservation and the EOS. They are pressure-state coefficients and remain distinct from any mechanical stress impulse.

\section{Finite-correlation work regularisation and reservoir}
\label{sec:reservoir}

Let $\E$ denote expectation and let $\xi^\tau$ be the stationary Ornstein--Uhlenbeck process
\begin{equation}
 \dd\xi^\tau=-\tau^{-1}\xi^\tau\dd t+\tau^{-1}\dd W_t,
 \qquad
 \E[\xi^\tau(t)\xi^\tau(t+s)]=\frac{1}{2\tau}e^{-|s|/\tau}.
 \label{eq:OU}
\end{equation}
Its time integral has unit two-sided Green--Kubo covariance.  Define
\begin{equation}
 Q^\tau=(\xi^\tau)^2-\frac{1}{2\tau},
 \qquad
 \eta^\tau=\sqrt{2\tau}\,Q^\tau.
 \label{eq:chaos}
\end{equation}
Then
\begin{equation}
 (\xi^\tau)^2=\frac{1}{2\tau}+\frac{1}{\sqrt{2\tau}}\eta^\tau.
 \label{eq:squaredecomp}
\end{equation}
Thus a fixed random--random power coefficient has a divergent zero-chaos mean and a non-tight centred square.  Wick centring alone does not produce a finite power limit.

The ordinary fast kinetic energy
\begin{equation}
 K_f^\tau=\frac12\rho|\balpha|^2(\xi^\tau)^2
 =\frac{\rho|\balpha|^2}{4\tau}+\frac12\rho|\balpha|^2Q^\tau
 \label{eq:fastK}
\end{equation}
shows the same $\tau^{-1}$ equilibrium divergence.

At finite correlation time, let $\bm X_t^\tau$ be the particle trajectory, $\bu^\tau$ its resolved drift and $\bz^\tau$ the ordinary unresolved velocity. The displacement is
\begin{equation}
 \dd\bm X_t^\tau=\bu^\tau(\bm X_t^\tau,t)\,\dd t+\bz^\tau(\bm X_t^\tau,t)\,\dd t,
 \qquad
 \int_0^t\bz^\tau(s)\,\dd s\Longrightarrow\balpha W_t.
 \label{eq:finitecorrmap}
\end{equation}
Here $\Longrightarrow$ denotes convergence in distribution on path space. A finite-correlation reservoir is therefore introduced in Hilbert spaces $\mathcal H_s$ and $\mathcal H_z$. Here $\bw\in\mathcal H_s$ is the resolved coupled state, $\bz^\tau\in\mathcal H_z$ is the unresolved state, and $\bm f_s,\bm f_z$ are specified inputs:
\begin{equation}
\begin{aligned}
 \dd\bw&=(-\mathsf R\bw+\mathsf C_\tau\bz^\tau+\bm f_s)\dd t,\\
 \dd\bz^\tau&=[-\mathsf C_\tau^*\bw-(\mathsf G_\tau+\mathsf J_\tau)\bz^\tau+\bm f_z]\dd t
 +\mathsf B_\tau\dd W_t,
\end{aligned}
\label{eq:reservoirSDE}
\end{equation}
where $\mathsf R=\mathsf R^*>0$ on the coupled resolved subspace, $\mathsf G_\tau=\mathsf G_\tau^*>0$ and $\mathsf J_\tau^*=-\mathsf J_\tau$. Inner products and norms carrying subscripts $s$ and $z$ are those of $\mathcal H_s$ and $\mathcal H_z$. Here $\mathsf B_\tau\in\mathcal L(\mathbb R,\mathcal H_z)$ and the same scalar $W_t$ is used in the one-channel finite-correlation precursor. The ordinary reservoir energy density $\cR_{\rm ord}^\tau(\bx)$ is chosen so that $\int_\Omega\cR_{\rm ord}^\tau\,\dd V=\|\bz^\tau\|_z^2/2$.  Adjoint coupling gives
\begin{equation}
 \langle\bw,\mathsf C_\tau\bz^\tau\rangle_s
 -\langle\bz^\tau,\mathsf C_\tau^*\bw\rangle_z=0,
 \label{eq:adjointcancel}
\end{equation}
so resolved--unresolved work is sign-neutral in the total energy.

The closure embedding is required to satisfy
\begin{equation}
 \int_\Omega\mathscr X_H\,\dd V
 =\langle\bw,\mathsf C_\tau\bz^\tau\rangle_s
 =\langle\bz^\tau,\mathsf C_\tau^*\bw\rangle_z
 \label{eq:HRembedding}
\end{equation}
in the H--R mechanical sector, up to explicitly specified reversible pressure and source-alignment exchanges. Equation \eqref{eq:HRembedding} is the bridge between the local split \eqref{eq:HRmatching} and the Hilbert-space reservoir. It is a constitutive matching condition, not a consequence of total-energy conservation alone.

The feedback-corrected calibration is
\begin{equation}
 \mathsf S_\tau=\mathsf G_\tau+\mathsf J_\tau+\mathsf C_\tau^*\mathsf R^{-1}\mathsf C_\tau,
 \qquad
 \mathsf B_\tau=\mathsf S_\tau\balpha.
 \label{eq:GKcal}
\end{equation}
Here $\mathsf R^{-1}$ is the inverse on the coupled resolved subspace (or the corresponding generalised inverse when null modes are projected out). It recovers
\begin{equation}
 \int_{-\infty}^{\infty}\E[\bz^\tau(t)\otimes\bz^\tau(0)]\,\dd t
 =\balpha\otimes\balpha.
 \label{eq:GK}
\end{equation}
The stationary covariance $\bm\Sigma_{zz}^\tau$ solves the corresponding Lyapunov equation, $\mathsf D_\tau=\mathsf B_\tau\mathsf B_\tau^*$ denotes the reservoir diffusion covariance, and $\mathsf M_z$ is the positive energy metric defining $\|\cdot\|_z$. Let the local equilibrium counterterm density be
\begin{equation}
 \cE_{\rm eq}^\tau(\bx)
 =\frac12\tr\left[\mathsf M_z(\bx)\bm\Sigma_{zz}^\tau(\bx,\bx)\right],
 \qquad
 E_{\rm eq}^\tau=\int_\Omega\cE_{\rm eq}^\tau\,\dd V
 =\frac12\operatorname{Tr}(\mathsf M_z\bm\Sigma_{zz}^\tau).
 \label{eq:eqcounterdensity}
\end{equation}
Let $U^\tau$ denote the finite-correlation physical internal-energy density. Define
\begin{equation}
 \widehat\cR^\tau=\cR_{\rm ord}^\tau-\cE_{\rm eq}^\tau,
 \qquad
 \widehat U^\tau=U^\tau+\cE_{\rm eq}^\tau.
 \label{eq:counterterm}
\end{equation}
The combined local and global energies are gauge invariant.  State-dependent metrics, projectors and calibration require the full It\^o differential of $\cE_{\rm eq}^\tau$, including Hessian and connection terms \citep{Hottovy2015,Birrell2017}.

Let $\mathsf G_{\rm visc}$ denote the positive Stokes-type relaxation operator on the unresolved velocity. Its random--random viscous quadratic form is realised once as unresolved relaxation and once, with opposite sign, as thermodynamic heating:
\begin{equation}
 \langle\bz^\tau,\mathsf G_{\rm visc}\bz^\tau\rangle
 =\int_\Omega[2\mu|\dev\bm h_1|^2+\zeta\theta_1^2](\xi^\tau)^2\,\dd V.
 \label{eq:fastvisc}
\end{equation}
For the augmented pressure carrier introduced in \cref{sec:pressurecarrier}, the unresolved equation includes the reciprocal force \cref{eq:pressureforce}. The pressure--volume exchange cancels between \cref{eq:pressurekinetic} and \cref{eq:pressureenergybalance}, leaving the boundary flux and relaxation loss in \cref{eq:pressurepairenergy}. No separate random--random pressure product is inserted into the bulk white-noise baseline.

The white-noise formulation retains one Brownian displacement channel.  The centred second-chaos contribution is an exact fast corrector that cancels inside the finite-$\tau$ combined energy before the limit; no independent $W^{(2)}$ energy channel is retained.

\section{Finite-correlation mechanical pressure carrier}
\label{sec:pressurecarrier}

The material pressure martingale coefficient $p_1^{\rm M}=\balpha\cdot\nabla p$ in \cref{eq:pMaterialMart} has dimensions $[p]\mathsf T^{-1/2}$. It is a pressure-state rate generated by transport through the equation of state, not a stress impulse: thermodynamic pressure enters the Cauchy stress as $-p\bI\,\dd t$, and $p_1^{\rm M}$ does not create a separate $\dd W_t$ stress measure. A direct ansatz $p_f^\tau=\Pi\xi^\tau$, where $p_f^\tau$ is a fast pressure, $\Pi$ its amplitude and $\xi^\tau$ a coloured carrier with correlation time $\tau$, would require density or temperature fluctuations with variance of order $\tau^{-1}$; the quadratic thermodynamic free energy would then diverge at the same order and would have to be included in an augmented reservoir. Thus a non-zero fast mechanical pressure cannot be obtained from the regular pressure-state It\^o differential alone.

To represent such a fast pressure explicitly, consider the material perturbation $\pi^\tau$ associated with the finite-correlation unresolved velocity $\bz^\tau$. For a frozen stable thermodynamic state, define the adiabatic bulk modulus
\begin{equation}
 K_s=\rho c_s^2,
 \qquad
 c_s^2=\left(\frac{\partial p}{\partial\rho}\right)_s>0.
 \label{eq:Ks}
\end{equation}
Let $\D_t^\star=\partial_t+\bu^\star\cdot\nabla$ denote the resolved material derivative under frozen slow coefficients. The reversible linearised equation of state gives $\D_t^\star\pi^\tau=-K_s\Div\bz^\tau$. A minimal causal relaxation completion is
\begin{equation}
 \D_t^\star\pi^\tau+\tau_p^{-1}\pi^\tau
 =-K_s\Div\bz^\tau,
 \qquad \tau_p>0.
 \label{eq:pressurecarrier}
\end{equation}
No independent Brownian forcing is added to \cref{eq:pressurecarrier}; the carrier is driven by the same finite-correlation velocity that produces the displacement channel. The unresolved mechanical equation is augmented by the reciprocal pressure force
\begin{equation}
 \left.\D_t^\star\bz^\tau\right|_p
 =-\rho^{-1}\nabla\pi^\tau.
 \label{eq:pressureforce}
\end{equation}
Equations \eqref{eq:pressurecarrier} and \eqref{eq:pressureforce} form an energy-conjugate acoustic pair. The ordinary quadratic pressure energy density is
\begin{equation}
 \cR_p^\tau=\frac{(\pi^\tau)^2}{2K_s}.
 \label{eq:pressureenergy}
\end{equation}
For frozen $K_s$, multiplication of \cref{eq:pressurecarrier} by $\pi^\tau/K_s$ yields
\begin{equation}
 \D_t^\star\cR_p^\tau
 =-\pi^\tau\Div\bz^\tau
 -\frac{(\pi^\tau)^2}{\tau_pK_s}.
 \label{eq:pressureenergybalance}
\end{equation}
The first term is the reversible pressure--volume exchange with the mechanical sector. Indeed, the pressure part of the unresolved kinetic-energy equation is
\begin{equation}
 \left.\D_t^\star\left(\frac12\rho|\bz^\tau|^2\right)\right|_p
 =-\Div(\pi^\tau\bz^\tau)+\pi^\tau\Div\bz^\tau,
 \label{eq:pressurekinetic}
\end{equation}
so the exchange cancels in the sum:
\begin{equation}
 \left.\D_t^\star\left(\frac12\rho|\bz^\tau|^2+\cR_p^\tau\right)\right|_p
 +\Div(\pi^\tau\bz^\tau)
 =-\frac{(\pi^\tau)^2}{\tau_pK_s}.
 \label{eq:pressurepairenergy}
\end{equation}
Here the energy densities are understood in the frozen-background fast linearisation; slow transport of the state-dependent metric is included with the connection and counterterm terms of the general reservoir. The relaxation term is non-negative heating when transferred to the physical internal energy, and contributes
\begin{equation}
 \sigma_p^\tau=\frac{(\pi^\tau)^2}{\tau_pK_sT}\ge0
 \label{eq:pressureentropy}
\end{equation}
to the finite-correlation entropy production. Thus a local algebraic identification of fast pressure with the divergence rate would hide both the carrier energy and its relaxation entropy.

\begin{figure}[ht]
 \centering
 \includegraphics[width=0.70\textwidth]{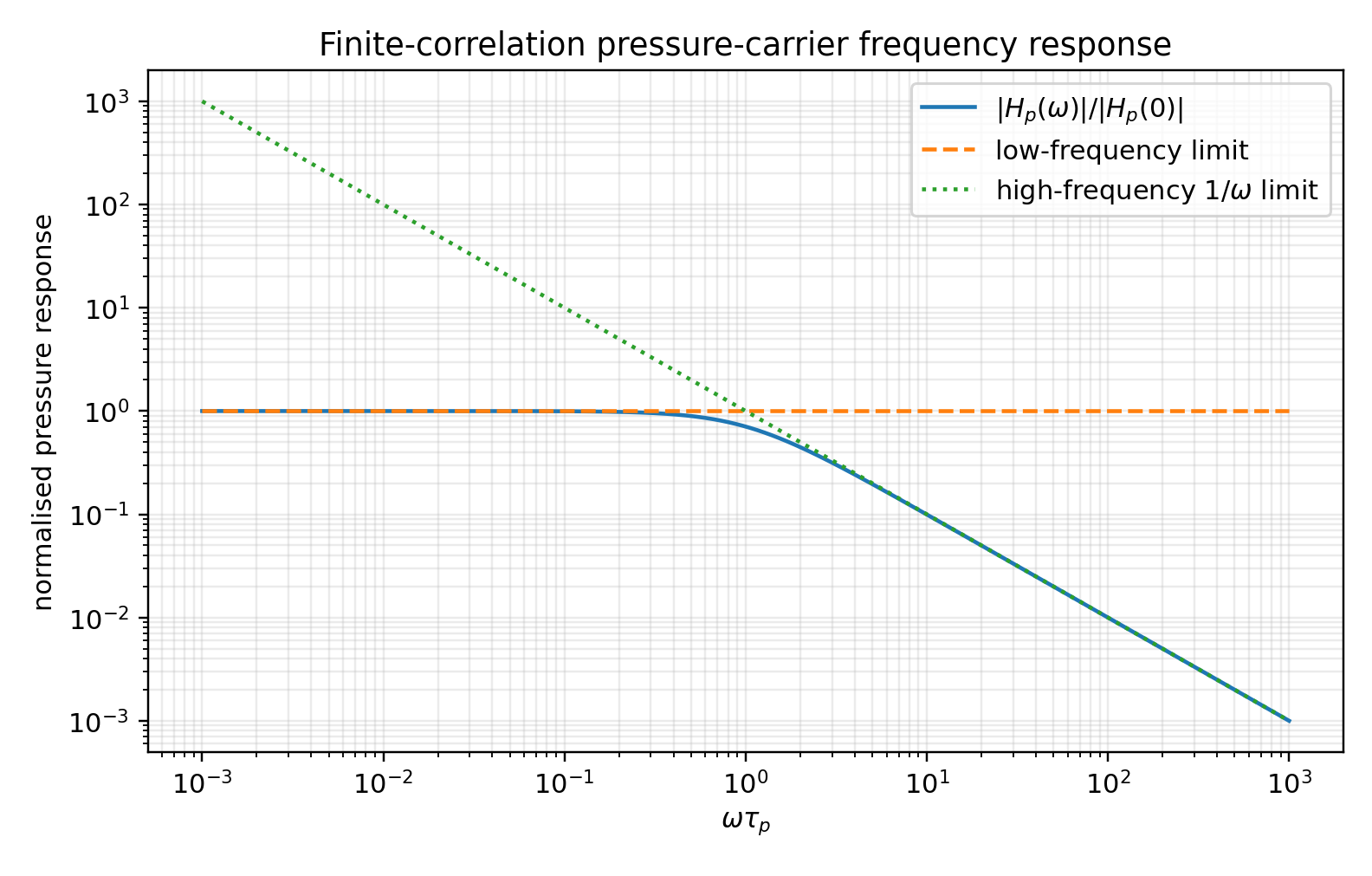}
 \caption{Frequency response of the finite-correlation pressure carrier. The response is normalised by its zero-frequency magnitude.}
 \label{fig:pressurecarrier}
\end{figure}

For a one-channel frozen mode $\bz^\tau=\balpha\xi^\tau$ with $\theta_1=\Div\balpha$, let $\omega$ be angular frequency and let hats denote temporal Fourier transforms. The frequency response is
\begin{equation}
 \widehat{\pi}^\tau(\omega)
 =-\frac{K_s\theta_1}{i\omega+\tau_p^{-1}}
 \widehat{\xi}^\tau(\omega).
 \label{eq:pressurefrequency}
\end{equation}
The normalised magnitude of this response is shown in \cref{fig:pressurecarrier}. The carrier follows the quasi-static bulk response for $|\omega|\tau_p\ll1$ and decays as $|\omega|^{-1}$ at high frequency.
At frequencies $|\omega|\tau_p\ll1$, the formal zero-frequency coefficient multiplying $\xi^\tau$ is
\begin{equation}
 \Pi_{1,\mathrm{eff}}=-K_s\tau_p\theta_1.
 \label{eq:Pieff}
\end{equation}
This quantity has the dimensions of a stress impulse, but \cref{eq:Pieff} is not a stand-alone white-noise constitutive law. Eliminating $\pi^\tau$ at low frequency converts its relaxation into an additional bulk-viscous response and discards the energy \cref{eq:pressureenergy}; it is admissible only when the carrier energy, counterterm and relaxation heating are retained in the finite-correlation reservoir.

This observation gives the regular finite-Mach closure used in the expanded white-noise governing equations: the martingale stress is the second line of \cref{eq:stress}, with no additional isotropic pressure-impulse term. This does not suppress thermodynamic pressure fluctuations, which are already contained in $p(\rho,T)$ and in the material It\^o coefficients \cref{eq:pMaterialCoefficients}. It states only that a regular equation-of-state fluctuation does not supply an additional distribution-valued bulk stress impulse. If fast mechanical pressure is required, \cref{eq:pressurecarrier} is retained as an augmented finite-correlation state rather than replaced by an unaccompanied $\Pi_{1,\mathrm{eff}}\dd W_t$ term.

When the carrier is retained, its energy $\cR_p^\tau$ is included in the finite-correlation reservoir $\cR_{\rm ord}^\tau$ and its relaxation loss is transferred once to physical internal energy. The carrier also separates three asymptotic regimes. With finite $K_s$ and $\tau_p$, it is an evolutionary acoustic-relaxation state. If $\tau_p\to0$ while $K_s\tau_p\to\zeta_p<\infty$, where $\zeta_p$ is the finite effective bulk-viscosity coefficient, then $\pi^\tau\to-\zeta_p\Div\bz^\tau$, which is an additional bulk-viscous closure. If $K_s\to\infty$ and $K_s\tau_p\to\infty$, the compliance and relaxation terms vanish relative to the pressure reaction, $\Div\bz^\tau=0$ becomes a constraint and pressure becomes a multiplier determined through the momentum equation. The last regime is the singular pathwise-isochoric limit associated with the incompressible theory; it is not a regular finite-Mach equation-of-state impulse.

\section{Stochastic Gibbs identity and entropy}
\label{sec:entropy}

For a simple substance, the material representation $e=e(s,v)$, $T=e_s>0$ and $p=-e_v$ is thermodynamically equivalent to the Helmholtz representation in \cref{eq:helmholtz}. Along the material path of \cref{eq:map}, write
\begin{equation}
 \dd e=e_0\dd t+e_1\dd W_t,
 \qquad
 \dd s=s_0\dd t+s_1\dd W_t.
\end{equation}
Subscripts $s$ and $v$ denote thermodynamic partial derivatives. Define the thermodynamic martingale coefficients
\begin{equation}
 T_1^{\rm th}=T_ss_1+T_vv\theta_1,
 \qquad
 p_1^{\rm M}=p_ss_1+p_vv\theta_1.
\end{equation}
It\^o's formula gives the exact Gibbs identity
\begin{equation}
T\dd s=\dd e+p\dd v
 -\frac12(T_1^{\rm th}s_1-p_1^{\rm M}v\theta_1)\dd t.
\label{eq:Gibbs}
\end{equation}
Equivalently,
\begin{equation}
 \rho Ts_1=\rho e_1+p\theta_1,
 \label{eq:Gibbsm}
\end{equation}
\begin{equation}
 \rho Ts_0=\rho e_0+p\Theta
 -\frac12(\rho T_1^{\rm th}s_1-p_1^{\rm M}\theta_1).
 \label{eq:Gibbsd}
\end{equation}
The half bracket is a reversible coordinate-change correction.  For isentropic acoustic compression, $s_1=0$ and
\begin{equation}
 e_1=-pv\theta_1,
 \qquad
 p_1^{\rm M}=-\rho c_s^2\theta_1,
 \qquad
 c_s^2=\left(\frac{\partial p}{\partial\rho}\right)_s.
 \label{eq:isentropicp}
\end{equation}
These equations are the material-path form of the same thermodynamic pressure state obtained from the equation of state; they do not define a mechanical stress impulse. The effective coefficient in \cref{eq:Pieff} belongs to the separately retained pressure carrier.

The physical internal energy is related to the centred reservoir gauge by
\begin{equation}
 e=\widehat e^\tau-\epsilon_{\rm eq}^\tau,
 \qquad
 \epsilon_{\rm eq}^\tau=\cE_{\rm eq}^\tau/\rho.
\end{equation}
Its Gibbs differential must include the complete It\^o differential of $\epsilon_{\rm eq}^\tau$.  Missing Hessian or moving-projector terms create spurious entropy sources.

The bulk heat-flux law is the deterministic Fourier choice \cref{eq:Fourier}. If a heat-flux impulse were introduced, its dimension would be $[\bq_1]=[\bq_0]\mathsf T^{1/2}$, not that of an ordinary heat flux. Such an extension must be derived from an augmented thermal conservation law rather than postulated independently.

For a finite-dimensional Galerkin truncation of the frozen unresolved Gaussian process, let $f_\tau$ be its probability density and let $\pi_\tau=\mathcal N(0,\bm\Sigma_{zz}^\tau)$ be the invariant Gaussian density. Define
\begin{equation}
 \mathscr H_z^\tau[f_\tau|\pi_\tau]
 =\int f_\tau\log(f_\tau/\pi_\tau)\,\dd\bz.
\end{equation}
The Fokker--Planck equation gives
\begin{equation}
 \frac{\dd}{\dd t}\mathscr H_z^\tau
 =-\frac12\mathscr I_{\mathsf D_\tau}\le0.
 \label{eq:relentropy}
\end{equation}
Here $\mathscr I_{\mathsf D_\tau}$ is the diffusion-weighted relative Fisher information. Under local detailed balance in the sense of thermodynamically consistent stochastic modelling \citep{Ottinger1998,Serrano2001}, $k_{\rm res}>0$ is the entropy scale of the reservoir (equal to the Boltzmann constant $k_B$ in a molecular thermal interpretation). The negative relative entropy is a free-entropy functional combining unresolved Shannon entropy and thermostat heat.  It must not be added a second time to the explicit thermostat entropy flux.

The entropy statement is conditional on the following closure hypotheses: (i) $T>0$ and a $C^2$ stable equation of state; (ii) $\mu,\zeta,\kappa\ge0$; (iii) the H--R exchange matching \eqref{eq:HRmatching}; (iv) kinetic-quadratic-variation and source-alignment terms are paired as reversible internal--reservoir exchanges unless a separate positive dissipation is specified; (v) the reservoir satisfies local detailed balance and the moving-equilibrium connection terms are retained; and (vi) $\bq_1=\bzero$ in the bulk formulation. Under these hypotheses, the predictable total entropy-production rate $\Sigma_{H-R}^\tau$ obeys
\begin{equation}
\begin{aligned}
\Sigma_{H-R}^\tau
={}&\int_\Omega\left[
 \frac{2\mu|\dev\bm h_0|^2+\zeta\Theta^2}{T}
 +\kappa\frac{|\nabla T|^2}{T^2}\right]\dd V
 +\frac{k_{\rm res}}{2}\mathscr I_{\mathsf D_\tau}\\
&+\int_\Omega\frac{(\pi^\tau)^2}{\tau_pK_sT}\,\dd V\ge0,
\end{aligned}
 \label{eq:entropyprod}
\end{equation}
where the last integral is included only when the augmented pressure carrier is retained; it vanishes in the regular finite-Mach white-noise baseline.
The total entropy balance is
\begin{equation}
 \dd\mathscr S_{\rm tot}^\tau
 =\Sigma_{H-R}^\tau\dd t
 +\dd\mathscr S_{\rm supply}
 +\dd\mathscr S_{\rm boundary}
 +\dd\mathscr M_S,
 \label{eq:entropybal}
\end{equation}
where $\mathscr S_{\rm supply}$ and $\mathscr S_{\rm boundary}$ are entropy supplied volumetrically and through the boundary, and $\mathscr M_S$ is an exchange martingale, not production.  For a closed adiabatic system with an integrable martingale,
\begin{equation}
 \frac{\dd}{\dd t}\E\mathscr S_{\rm tot}^\tau
 =\E\Sigma_{H-R}^\tau\ge0.
\end{equation}

\section{Entropy and finite-correlation reservoir boundary conditions}
\label{sec:entropybc}

The boundary conditions in \cref{sec:bc} close the primary mass, momentum and total-energy balances. The entropy and finite-correlation reservoir require the additional conditions stated here.

\subsection{Entropy flux and thermal entropy supply}

The deterministic entropy supply through heat is
\begin{equation}
 \dot{\mathscr S}_{\Gamma,q}=-\int_\Gamma\frac{\bq_0\cdot\bn}{T}\,\dd S,
 \label{eq:entropyheatport}
\end{equation}
with the sign defined by the outward normal. Mechanical wall work and mass-carried entropy must be added on moving or open boundaries.

For $\bq_1=\bzero$, the fixed-boundary entropy fluxes are
\begin{equation}
 \bm\cF_S^0=\rho s\bu_c+\balpha(\rho s_1)-\frac12\bSigmaAlpha\nabla(\rho s)+\frac{\bq_0}{T},
 \qquad
 \bm\cF_S^1=\rho s\balpha,
 \label{eq:entropyflux}
\end{equation}
up to specified volumetric entropy supplies. On a deterministic moving boundary, subtract $\rho s\bw_\Gamma$ from the drift flux. These total fluxes, rather than the conductive term alone, determine the entropy transfer at an open boundary.

\subsection{Finite-correlation reservoir boundary}

A closed bulk reservoir may use an energy-neutral/no-flux boundary operator. For a dynamic passive wall, $\mathsf T$ is the trace map, $\bm g$ the wall state, $\dd\bm\Lambda_\Gamma$ the constraint reaction, $\mathsf L$ a lifting operator, $\mathsf M_\Gamma$ a positive wall metric, $\mathsf Z_\Gamma$ the wall impedance, $\mathsf B_\Gamma$ the wall-noise amplitude and $\bm W_t^\Gamma$ a boundary Wiener process. The coupled law is
\begin{equation}
\begin{aligned}
 \dd\bz^\tau&=\cdots+\mathsf L\dd\bm\Lambda_\Gamma,\\
 \mathsf M_\Gamma\dd\bm g
 &=-\mathsf Z_\Gamma\bm g\dd t+\mathsf B_\Gamma\dd\bm W_t^\Gamma-\dd\bm\Lambda_\Gamma,\\
 \mathsf T\bz^\tau&=\bm g.
\end{aligned}
\label{eq:dynamicwall}
\end{equation}
The fluid reaction work $(\mathsf T\bz^\tau)\cdot\dd\bm\Lambda_\Gamma$ cancels the wall reaction work $-\bm g\cdot\dd\bm\Lambda_\Gamma$ on the constraint. A passive wall requires $\mathsf Z_\Gamma\ge0$ and a fluctuation--dissipation-compatible $\mathsf B_\Gamma$; an active wall may instead be calibrated to a target displacement covariance but represents external energy input \citep{Tsai2026PaperI}.

\subsection{Closed adiabatic system}

Combining the primary conditions \cref{eq:closedmassenergy} with the entropy flux gives the closed adiabatic conditions
\begin{equation}
 \bm\cF_\rho^0\cdot\bn=0,
 \qquad
 \bm\cF_\rho^1\cdot\bn=0,
 \qquad
 \bm\cF_E^0\cdot\bn=0,
 \qquad
 \bm\cF_E^1\cdot\bn=0,
 \qquad
 \bm\cF_S^0\cdot\bn=0,
 \qquad
 \bm\cF_S^1\cdot\bn=0,
 \label{eq:closedadiabatic}
\end{equation}
together with an energy-neutral or detailed-balance reservoir boundary and an integrable entropy-exchange martingale. On a deterministic moving boundary, use the relative drift fluxes in \cref{eq:movingflux}; non-zero mechanical wall work is external energy input rather than internal generation.

A simple sufficient fixed-wall specialisation is $\bu=\bzero$, $\balpha=\bzero$, $\bq_0\cdot\bn=0$, and a no-flux reservoir boundary. If tangential volatility is retained, the complete martingale energy flux must still vanish. In particular, a condition involving only $(\bm\cF_m^1\bn)\cdot\bu^\star$ is insufficient because $(\bT_0\bn)\cdot\balpha$ remains in $\bm\cF_E^1\cdot\bn$.

Under \cref{eq:closedadiabatic}, absent body power and volumetric heating, global total energy is conserved and expected total entropy is non-decreasing.

\section{Ideal-gas specialization of the governing system}
\label{sec:idealgas}

For a calorically perfect ideal gas, let
\begin{equation}
 p=\rho R_{\rm g}T,
 \qquad
 e=c_vT,
 \qquad
 c_p=c_v+R_{\rm g},
 \qquad
 \gamma=\frac{c_p}{c_v},
 \qquad
 c_s^2=\gamma R_{\rm g}T.
 \label{eq:idealthermo}
\end{equation}
Here $R_{\rm g}$ is the specific gas constant, $c_v$ and $c_p$ are the constant-volume and constant-pressure specific heats, $\gamma$ is their ratio, and $c_s$ is the adiabatic sound speed. For direct implementation, the transport and kinematic quantities used throughout this section are restated here:
\begin{equation}
 \bmom=\rho\bu,
 \qquad
 \bSigmaAlpha=\balpha\otimes\balpha,
 \qquad
 \bu_c=\bu-\frac12\Div\bSigmaAlpha,
 \qquad
 \bu^\star=\bu_c+\balpha(\Div\balpha),
 \label{eq:idealimplementationdrifts}
\end{equation}
with
\begin{equation}
 \bA=\nabla\balpha,
 \qquad
 \bD_u=\sym(\nabla\bu),
 \qquad
 \bD_\alpha=\sym\bA,
 \qquad
 \bm h_1=\bD_\alpha,
 \qquad
 \bm h_0=\bD_u+\frac12\bA\bA^{\mathsf T}-\bD_\alpha^2,
 \label{eq:idealimplementationhencky}
\end{equation}
and
\begin{equation}
 \theta_1=\Div\balpha,
 \qquad
 \Theta=\Div\bu^\star.
 \label{eq:idealimplementationvolume}
\end{equation}
These definitions are repeated so that the ideal-gas subsystem can be read and implemented without returning to the kinematics section. The total-energy density becomes
\begin{equation}
 \cE=\rho c_vT+\cR+\frac12\rho|\bu|^2+\rho\Phi.
 \label{eq:idealE}
\end{equation}
Define the viscous parts
\begin{equation}
 \bm\tau_0=2\mu\dev\bm h_0+\zeta\Theta\bI,
 \qquad
 \bm\tau_1=2\mu\dev\bm h_1+\zeta\theta_1\bI.
\end{equation}
Then
\begin{equation}
 \bT_0=-\rho R_{\rm g}T\bI+\bm\tau_0,
 \qquad
 \bT_1=\bm\tau_1,
 \label{eq:idealstress}
\end{equation}
and $\bq_0=-\kappa\nabla T$, $\bq_1=\bzero$.

For clarity, all ideal-gas channels are written explicitly as evolution partial differential equations and martingale compatibility equations. The mass equations are
\begin{subequations}
\label{eq:idealmass}
\begin{align}
 \partial_t\rho+\Div(\rho\bu_c)
 &=\frac12\Div(\bSigmaAlpha\nabla\rho),
 \label{eq:idealmassdrift}\\
 \Div(\rho\balpha)&=0.
 \label{eq:idealmassmart}
\end{align}
\end{subequations}
The momentum equations are
\begin{subequations}
\label{eq:idealmomentum}
\begin{align}
 \partial_t(\rho\bu)+\Div(\rho\bu\otimes\bu_c)
 +\Div(\bQ_1\otimes\balpha)
 &=\frac12\Div(\bSigmaAlpha\nabla(\rho\bu))
 +\Div(-\rho R_{\rm g}T\bI+\bm\tau_0)+\rho\bm b_0,
 \label{eq:idealmomentumdrift}\\
 \Div(\rho\bu\otimes\balpha)
 &=\Div\bm\tau_1+\rho\bm b_1.
 \label{eq:idealmomentummart}
\end{align}
\end{subequations}
Here $\bQ_1=\Div\bm\tau_1+\rho\bm b_1$ in the regular finite-Mach white-noise baseline.

The energy-source amplitudes specialize to
\begin{equation}
\begin{aligned}
 R_{0,\rm ig}^E={}&\Div[(-\rho R_{\rm g}T\bI+\bm\tau_0)\bu^\star]
 +\Div(\kappa\nabla T)
 +\rho\bm b_0^{nc}\cdot\bu^\star+r_0,\\
 R_{1,\rm ig}^E={}&\Div[(-\rho R_{\rm g}T\bI+\bm\tau_0)\balpha
 +\bm\tau_1\bu^\star]\\
 &+\rho\bm b_0^{nc}\cdot\balpha
 +\rho\bm b_1^{nc}\cdot\bu^\star+r_1.
\end{aligned}
\label{eq:idealEsources}
\end{equation}
Hence the total-energy equations are
\begin{subequations}
\label{eq:idealenergy}
\begin{align}
 \partial_t\cE+\Div(\cE\bu_c)+\Div(\balpha R_{1,\rm ig}^E)
 &=\frac12\Div(\bSigmaAlpha\nabla\cE)+R_{0,\rm ig}^E,
 \label{eq:idealenergydrift}\\
 \Div(\cE\balpha)&=R_{1,\rm ig}^E.
 \label{eq:idealenergymart}
\end{align}
\end{subequations}

For $\varepsilon=\rho c_vT$, the physical internal-energy equations \cref{eq:physicalinternalchannels} become the temperature pair
\begin{subequations}
\label{eq:idealT}
\begin{align}
 \rho c_v\left(\partial_tT+\bu_c\cdot\nabla T\right)
 &=\frac{c_v}{2}\left[\Div(\rho\bSigmaAlpha\nabla T)
 +(\bSigmaAlpha\nabla\rho)\cdot\nabla T\right]
 +\mathcal H_0-\Div(\balpha\mathcal H_1),
 \label{eq:idealTdrift}\\
 \rho c_v\,\balpha\cdot\nabla T&=\mathcal H_1.
 \label{eq:idealTmart}
\end{align}
\end{subequations}
The second line is the temperature martingale compatibility relation; it gives the material temperature martingale coefficient in \cref{eq:materialTIto}.

Using $p=\rho R_{\rm g}T$, \cref{eq:idealmass,eq:idealT} give the pressure pair
\begin{subequations}
\label{eq:idealpSDE}
\begin{align}
 (\partial_t+\bu_c\cdot\nabla)p+p\Div\bu_c
 &=\frac12\Div(\bSigmaAlpha\nabla p)
 +(\gamma-1)\left[\mathcal H_0-\Div(\balpha\mathcal H_1)\right],
 \label{eq:idealpdrift}\\
 \Div(p\balpha)&=(\gamma-1)\mathcal H_1.
 \label{eq:idealpmart}
\end{align}
\end{subequations}
Equivalently, the EOS-derived material pressure martingale coefficient is
\begin{equation}
 p_1^{\rm M}=\balpha\cdot\nabla p
 =-p\theta_1+(\gamma-1)\mathcal H_1.
 \label{eq:idealpMaterialMart}
\end{equation}
No independent Eulerian pressure martingale has been postulated: \cref{eq:idealpmart,eq:idealpMaterialMart} follow from the mass and physical internal-energy compatibility equations.

For the optional finite-correlation mechanical pressure carrier, $K_s=\rho c_s^2=\gamma p$, and \cref{eq:pressurecarrier} becomes
\begin{equation}
 \D_t^\star\pi^\tau+\tau_p^{-1}\pi^\tau
 =-\gamma p\,\Div\bz^\tau,
 \qquad
 \cR_p^\tau=\frac{(\pi^\tau)^2}{2\gamma p}.
 \label{eq:idealpressurecarrier}
\end{equation}
The white-noise ideal-gas equations above use the regular bulk closure in \cref{eq:stress}; \cref{eq:idealpressurecarrier} is retained only when a finite-correlation pressure-relaxation sector is explicitly resolved.

\subsection{Classical ideal-gas Navier--Stokes--Fourier limit}
\label{sec:idealNSFlimit}

The constitutive model is required to satisfy
\begin{equation}
 \balpha=\bzero
 \quad\Longrightarrow\quad
 \bm b_1=\bzero,\quad r_1=0,\quad \cR=0,
 \quad R_1^E=0,\quad \mathcal H_1=0,
 \label{eq:alphazero}
\end{equation}
unless an external stochastic forcing is deliberately retained. Then $\bSigmaAlpha=\bzero$, $\bu_c=\bu^\star=\bu$, $\bm h_0=\bD_u$, $\bm h_1=\bzero$ and $\Theta=\Div\bu$. With $\Phi=0$ for notational simplicity, the governing system becomes
\begin{align}
 \partial_t\rho+\Div(\rho\bu)&=0,
 \label{eq:NSFmass}\\
 \partial_t(\rho\bu)+\Div(\rho\bu\otimes\bu+p\bI-\bm\tau)
 &=\rho\bm b_0,
 \label{eq:NSFmomentum}\\
 \partial_t E_{\rm NSF}
 +\Div\left[(E_{\rm NSF}+p)\bu-\bm\tau\bu-\kappa\nabla T\right]
 &=\rho\bm b_0\cdot\bu+r_0,
 \label{eq:NSFenergy}
\end{align}
where
\begin{equation}
 E_{\rm NSF}=\rho c_vT+\frac12\rho|\bu|^2,
 \qquad
 \bm\tau=2\mu\dev\bD_u+\zeta(\Div\bu)\bI.
\end{equation}
Equivalently, the temperature and pressure equations are
\begin{equation}
 \rho c_v(\partial_tT+\bu\cdot\nabla T)
 =-p\Div\bu+\bm\tau:\nabla\bu
 +\Div(\kappa\nabla T)+r_0,
 \label{eq:NSFtemperature}
\end{equation}
\begin{equation}
 (\partial_t+\bu\cdot\nabla)p+\gamma p\Div\bu
 =(\gamma-1)\left[\bm\tau:\nabla\bu
 +\Div(\kappa\nabla T)+r_0\right].
 \label{eq:NSFpressure}
\end{equation}
Thus the zero-volatility constitutive limit recovers the standard calorically perfect ideal-gas compressible Navier--Stokes--Fourier equations, not merely their linear acoustic approximation. The boundary fluxes in \cref{sec:bc} reduce at the same time to the standard deterministic mass, traction and total-energy fluxes.

\section{Principal-symbol structure, directional degeneracy and singular limits}
\label{sec:type}

The primary unknowns in the present formulation are finite-variation Eulerian coefficient fields. Their drift equations form an evolutionary compressible system, while their Brownian channels are spatial compatibility constraints. The complete linearised object is therefore a descriptor system, not an unconstrained transport SPDE. At finite Mach number the drift block has acoustic and convective first-order branches together with viscous, thermal and volatility-induced second-order terms. It is mixed hyperbolic--parabolic and only partially parabolic. Elliptic operators enter through elimination of an algebraic solved-volatility block or through the singular low-Mach constraint limit.

\subsection{Frozen drift symbol and martingale constraint}

Consider a calorically perfect ideal gas linearised about a uniform rest state $(\rho_0,T_0,\bu=\bzero)$ with a prescribed constant base volatility $\balpha_0$. For a Fourier mode with wave vector $\bm k=k\bn$, define
\begin{equation}
 \delta=\rho'/\rho_0,
 \qquad
 \vartheta=T'/T_0,
 \qquad
 \beta=\balpha_0\cdot\bm k.
 \label{eq:typevariables}
\end{equation}
Let $\widehat{\bm y}=(\widehat\delta,\widehat u_\parallel,\widehat\vartheta)^{\mathsf T}$ and let $\widehat{\bm z}$ collect the linearised volatility, stress-impulse, reservoir and other algebraic closure variables. The frozen coefficient system has the descriptor form
\begin{subequations}
\label{eq:typeDescriptor}
\begin{align}
 \partial_t\widehat{\bm y}
 &=\left[\mathsf L_{NSF}(k)-\frac12\beta^2\mathsf I\right]
 \widehat{\bm y}+\mathsf C_0(\bm k)\widehat{\bm z},
 \label{eq:typeDriftSymbol}\\
 \bzero&=-i\beta\widehat{\bm y}+\mathsf C_1(\bm k)\widehat{\bm z}.
 \label{eq:typeMartSymbol}
\end{align}
\end{subequations}
Here the first line is the drift evolution symbol and the second is the martingale compatibility symbol. The deterministic Navier--Stokes--Fourier block is
\begin{equation}
 \mathsf L_{NSF}(k)=
 \begin{pmatrix}
  0 & -ik & 0\\
  -ikc_T^2 & -\nu_Lk^2 & -ikc_T^2\\
  0 & -ik(\gamma-1) & -\chi k^2
 \end{pmatrix},
 \label{eq:typeLNSF}
\end{equation}
with $c_T^2=R_{\rm g}T_0$, $c_s^2=\gamma c_T^2$, $\nu_L=(\zeta+4\mu/3)/\rho_0$ and $\chi=\kappa/(\rho_0c_v)$.

The term $-\beta^2\mathsf I/2$ is the directional second-order contribution visible in the drift equations before the martingale constraint is imposed. It cannot be interpreted by itself as an autonomous mean-diffusion operator. For the source-free passive comparison $\mathsf C_1\widehat{\bm z}=0$, \cref{eq:typeMartSymbol} requires
\begin{equation}
 \beta\widehat{\bm y}=0.
 \label{eq:sourcefreeMartSymbol}
\end{equation}
Thus non-trivial source-free modes are transverse to the base volatility, $\beta=0$, and the apparent volatility diffusion vanishes on the constraint manifold. Modes with $\beta\ne0$ require a compensating martingale stress, body-force, energy or solved-volatility source. Their admissible dynamics can be classified only after the corresponding closure block $\mathsf C_1$ has been specified.

When $\mu=\zeta=\kappa=0$, the unconstrained drift block has eigenvalues $0$ and $\pm ic_sk$. Viscosity and heat conduction add $O(k^2)$ damping but do not make the full state uniformly parabolic: density has no independent molecular diffusion in the classical NSF limit. The resolved drift subsystem is consequently partially dissipative and hyperbolic--parabolic, while the complete solved-volatility problem is differential--algebraic.

\subsection{Directional rank of the one-channel covariance}

The directional factor in \cref{eq:typeDescriptor} is
\begin{equation}
 \beta^2=\bm k^{\mathsf T}\bSigmaAlphaZero\bm k
 =|\balpha_0|^2k^2\cos^2\theta,
 \qquad
 \bSigmaAlphaZero=\balpha_0\otimes\balpha_0,
 \label{eq:betadirectional}
\end{equation}
where $\theta$ is the angle between $\bm k$ and $\balpha_0$. For one Brownian channel, $\bSigmaAlphaZero$ has rank at most one. A uniformly positive covariance in spatial dimension $d>1$ would require multiple channels,
\begin{equation}
 \bSigmaAlphaZero=\sum_{r=1}^{N_W}\balpha_{0r}\otimes\balpha_{0r}
 \succeq a_{\min}\mathsf I,
 \qquad a_{\min}>0.
 \label{eq:multichannelcoercivity}
\end{equation}
The rank statement concerns the transport covariance. The actual mode set is further restricted by the martingale compatibility block \cref{eq:typeMartSymbol}.

\subsection{Elliptic algebraic blocks and the low-Mach limit}

At finite Mach number thermodynamic pressure is obtained from the EOS and propagates through acoustic branches; it is not an elliptic Lagrange multiplier. An elliptic operator can nevertheless arise after a specific solved-volatility closure is supplied. A frozen descriptor system may be written as
\begin{equation}
 \mathsf M\,\partial_t\widehat{\bm y}
 =\mathsf L_e(\bm k)\widehat{\bm y}
 +\mathsf C(\bm k)\widehat{\bm z},
 \qquad
 \bzero=\mathsf H(\bm k)\widehat{\bm y}
 +\mathsf K(\bm k)\widehat{\bm z},
 \label{eq:descriptorSymbol}
\end{equation}
where $\mathsf M$ is the descriptor mass matrix, $\mathsf L_e$ the drift-evolution symbol, $\mathsf C$ and $\mathsf H$ coupling symbols, $\mathsf K$ the algebraic closure block, and $\widehat{\bm z}$ the algebraic variables. If $\mathsf K(\bm k)$ is invertible on the constrained subspace, then
\begin{equation}
 \widehat{\bm z}=-\mathsf K(\bm k)^{-1}\mathsf H(\bm k)\widehat{\bm y},
 \qquad
 \mathsf L_{\rm eff}=\mathsf L_e-\mathsf C\mathsf K^{-1}\mathsf H.
 \label{eq:schurSymbol}
\end{equation}
Ellipticity is a property of $\mathsf K(\bm k)$ and the associated boundary conditions, not of the conservation laws alone. The finite-correlation pressure carrier derived above is evolutionary and is not part of this algebraic block. The coercivity and index of the remaining solved-volatility block are closure-dependent.

The low-Mach limit is different. Let $Ma$ denote the Mach number. Acoustic eigenvalues scale as $\pm ik/Ma$ and become singular. For well-prepared data the fast acoustic component is filtered, density approaches its constraint manifold and the mechanical pressure becomes a multiplier. Taking the divergence of the limiting momentum equation together with the limiting mass constraint produces a Poisson- or Stokes-type elliptic pressure problem. This is a singular change of role as $Ma\to0$, not a smooth finite-Mach switch of the time-dependent PDE from parabolic to elliptic type.

\subsection{Pressure-carrier relaxation and limiting operator type}

The augmented pressure carrier \cref{eq:pressurecarrier} adds a longitudinal relaxation state. For frozen coefficients, its transfer function is \cref{eq:pressurefrequency}; the additional eigenvalue is $-\tau_p^{-1}$ before coupling to the acoustic pair. The reversible pressure--volume coupling is energy-skew with respect to the carrier metric $1/K_s$, while relaxation contributes the non-negative rate in \cref{eq:pressureentropy}. Consequently the finite-$K_s$, finite-$\tau_p$ system remains evolutionary and hyperbolic--parabolic--relaxational; it does not introduce an elliptic equation.

Two singular eliminations have different meanings. The scaling $\tau_p\to0$ with $K_s\tau_p\to\zeta_p$ produces a local bulk-viscous term and strengthens parabolic damping. The scaling $K_s\to\infty$, $K_s\tau_p\to\infty$ removes both compliance and finite relaxation from the leading-order volumetric balance, leaving a constraint whose multiplier is obtained through an elliptic Schur complement. Thus the pressure-carrier model gives an explicit bridge between acoustic relaxation, bulk-viscous reduction and the incompressible elliptic pressure limit.

\subsection{Admissible classification domain and numerical implications}

The local frozen-state classification requires
\begin{equation}
 \rho_0>0,
 \qquad T_0>0,
 \qquad c_v>0,
 \qquad c_s^2=\left(\frac{\partial p}{\partial\rho}\right)_s>0,
 \qquad \mu\ge0,
 \qquad \zeta\ge0,
 \qquad \kappa\ge0.
 \label{eq:typeadmissible}
\end{equation}
If $c_s^2$ approaches zero, acoustic hyperbolicity degenerates; if it becomes negative, the homogeneous thermodynamic state is unstable and a single-phase ideal-gas classification is no longer appropriate. If $\mu$ or $\kappa$ vanishes, the corresponding parabolic branch disappears. If the rank or orientation of $\bSigmaAlpha$ varies, the directional drift symbol and the martingale compatibility block both change locally.

Numerical validation should therefore monitor, in addition to conservation and entropy, the acoustic and thermal eigenvalues, the residual of every martingale compatibility equation, the directional quantity $\bm k^{\mathsf T}\bSigmaAlpha\bm k$, and the coercivity of any algebraic Schur complement. Subsonic, low-Mach, transonic and shock-containing calculations require different characteristic boundary treatments, but they do not reduce the complete descriptor system to a simple elliptic/parabolic switch.

\section{Consistency with canonical limits}
\label{sec:tests}

\subsection{Coefficient-level descriptor-symbol check}

For the frozen ideal-gas system \cref{eq:typeDescriptor}, the source-free martingale block is $-i\beta\widehat{\bm y}=0$. The symbolic check confirms that transverse modes with $\beta=0$ recover the deterministic NSF drift matrix \cref{eq:typeLNSF}, while non-trivial modes with $\beta\ne0$ require a compensating algebraic source. It also confirms that a one-channel covariance has rank one in three dimensions and that three independent spanning channels give a positive-definite covariance in the corresponding comparison calculation. This calculation checks the descriptor structure; it does not interpret the off-shell term $-\beta^2\mathsf I/2$ as an independent stochastic mean-diffusion law.

\subsection{Deterministic compressible Navier--Stokes--Fourier limit}

The complete calorically perfect ideal-gas reduction is given in \cref{sec:idealNSFlimit}. Under the zero-volatility constitutive condition \cref{eq:alphazero}, \cref{eq:mass,eq:momentum,eq:Econservative} become \cref{eq:NSFmass,eq:NSFmomentum,eq:NSFenergy}. The associated entropy production is
\begin{equation}
 \sigma_{NSF}=\frac{2\mu|\dev\bD_u|^2+\zeta(\Div\bu)^2}{T}
 +\kappa\frac{|\nabla T|^2}{T^2}\ge0.
\end{equation}

\subsection{Ideal-gas linear acoustic limit}

Linearise the deterministic system about $(\rho_0,T_0,\bu=\bzero)$ and consider a one-dimensional Fourier mode $e^{\lambda t+ikx}$.  With
\begin{equation}
 \delta=\rho'/\rho_0,
 \qquad
 \vartheta=T'/T_0,
 \qquad
 c_T^2=R_{\rm g}T_0,
 \qquad
 c_s^2=\gamma c_T^2,
\end{equation}
where $c_T$ and $c_s$ are the isothermal and adiabatic sound speeds. Define
\begin{equation}
 \nu_L=\frac{\zeta+4\mu/3}{\rho_0},
 \qquad
 \chi=\frac{\kappa}{\rho_0c_v},
\end{equation}
where $\nu_L$ is the longitudinal kinematic viscosity and $\chi$ the thermal diffusivity. The modal system is
\begin{equation}
\begin{aligned}
 \lambda\delta+iku&=0,\\
 (\lambda+\nu_Lk^2)u+ikc_T^2(\delta+\vartheta)&=0,\\
 (\lambda+\chi k^2)\vartheta+ik(\gamma-1)u&=0.
\end{aligned}
\label{eq:acousticmatrix}
\end{equation}
The exact dispersion polynomial is
\begin{equation}
\lambda^3+(\nu_L+\chi)k^2\lambda^2
 +(c_s^2k^2+\nu_L\chi k^4)\lambda
 +c_T^2\chi k^4=0.
\label{eq:acousticcubic}
\end{equation}
At small $k$,
\begin{equation}
 \lambda_\pm=\pm ic_sk-\Gamma k^2+O(k^3),
 \qquad
 \Gamma=\frac12\left[\nu_L+\frac{\gamma-1}{\gamma}\chi\right],
 \label{eq:soundasym}
\end{equation}
while
\begin{equation}
 \lambda_T=-\frac\chi\gamma k^2+O(k^4).
 \label{eq:thermalasym}
\end{equation}
The numerical roots agree with these expansions to better than $10^{-7}$ relative error over the specified low-$k$ interval.

\begin{figure}[ht]
 \centering
 \includegraphics[width=0.76\textwidth]{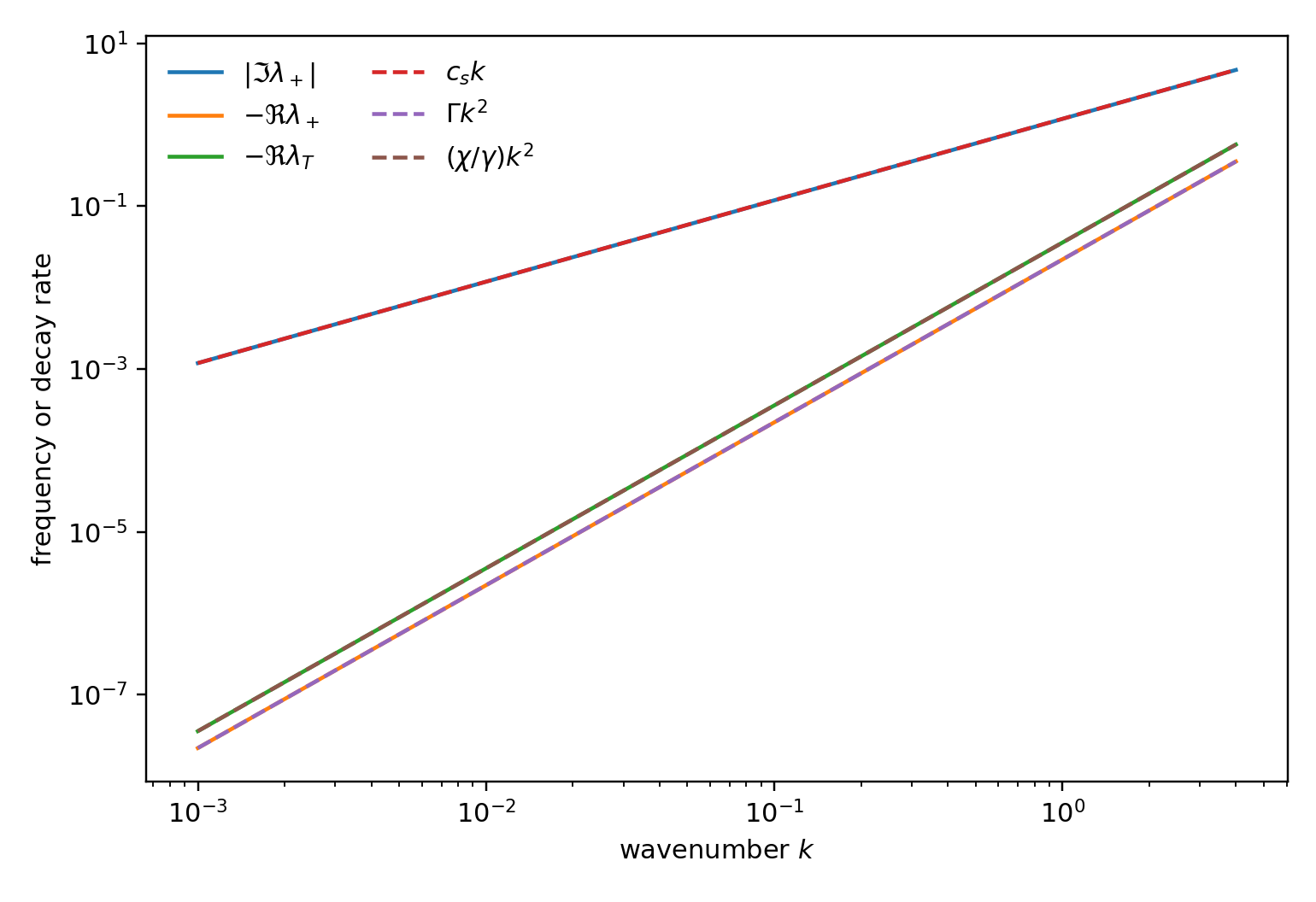}
 \caption{Ideal-gas linear acoustic calculation.  The exact cubic recovers the sound frequency, viscous--thermal attenuation and thermal diffusion branch.}
 \label{fig:dispersion}
\end{figure}

\subsection{One-dimensional viscous--thermal energy balance}

For a Fourier amplitude $\bm y=(\delta,u,\vartheta)$, a symmetrising quadratic energy is
\begin{equation}
 \mathscr E_2=\frac{\rho_0}{4}\left[c_T^2|\delta|^2+|u|^2+\frac{c_T^2}{\gamma-1}|\vartheta|^2\right].
 \label{eq:E2}
\end{equation}
It satisfies
\begin{equation}
 \frac{\dd\mathscr E_2}{\dd t}
 =-\frac{\rho_0}{2}\left[\nu_Lk^2|u|^2
 +\frac{c_T^2}{\gamma-1}\chi k^2|\vartheta|^2\right]\le0.
 \label{eq:E2diss}
\end{equation}
The matrix-exponential solution has a maximum integrated energy-balance error of $1.29\times10^{-5}$ relative to the initial energy and no monotonicity violation.

\begin{figure}[ht]
 \centering
 \includegraphics[width=0.48\textwidth]{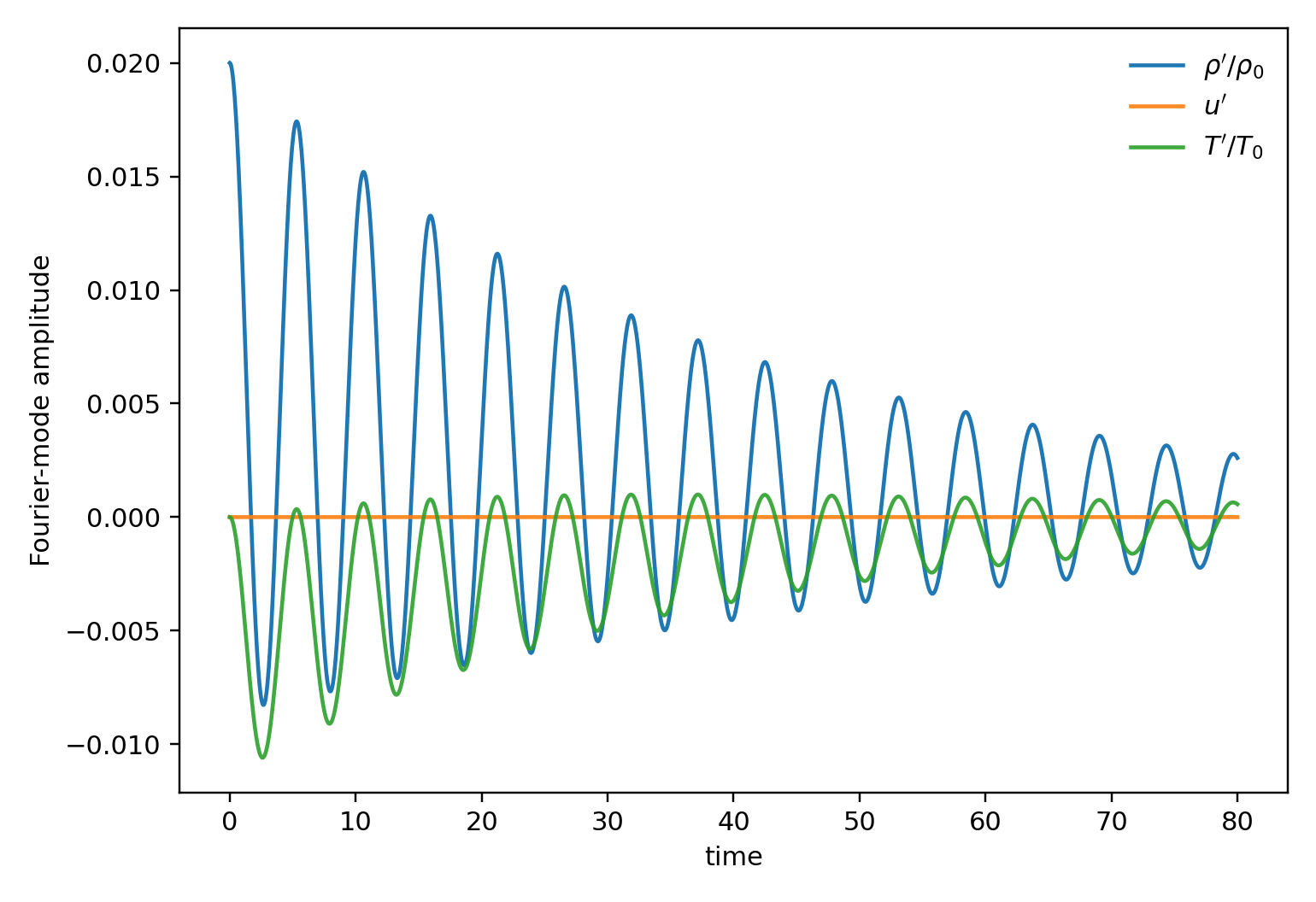}\hfill
 \includegraphics[width=0.48\textwidth]{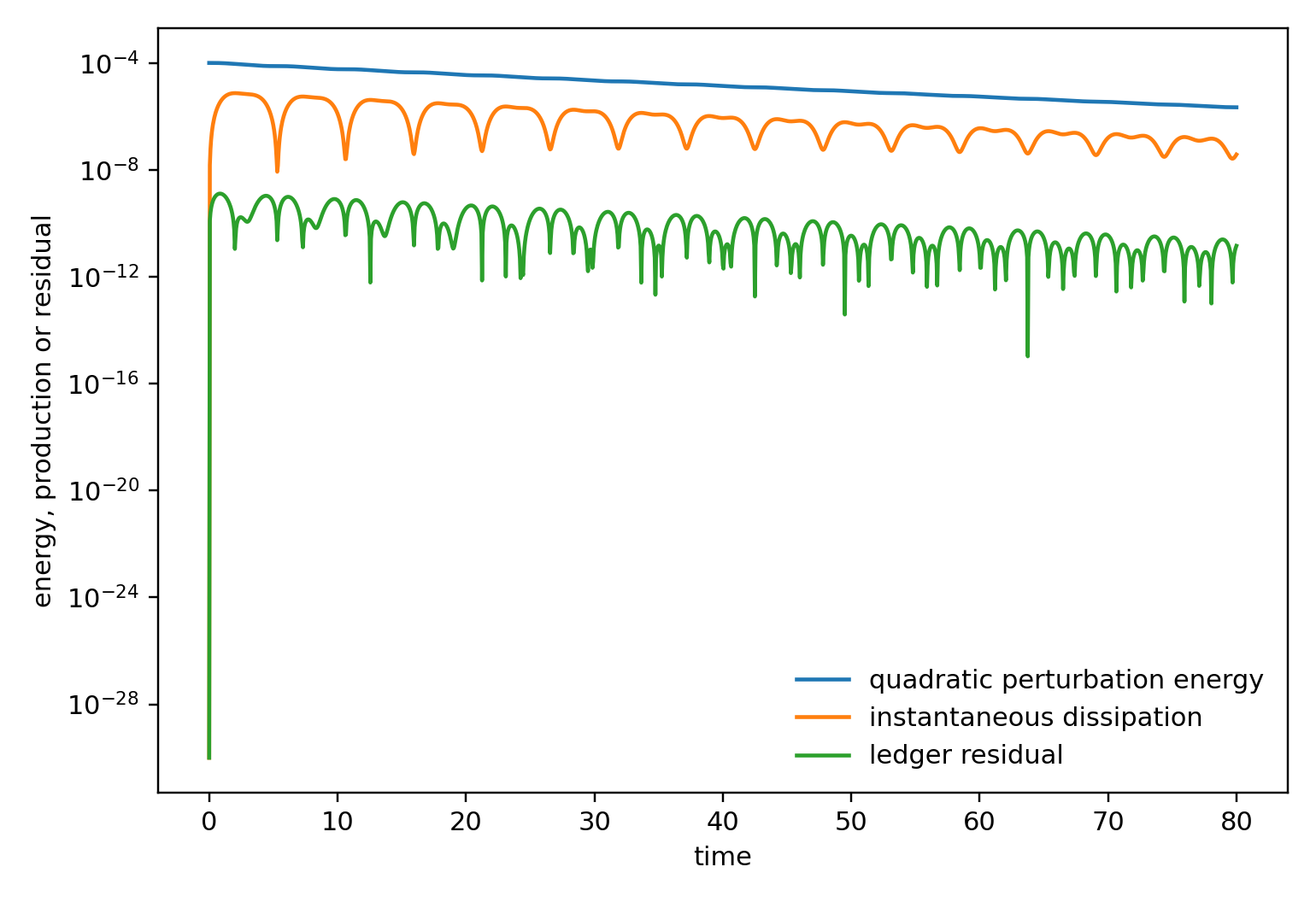}
 \caption{One-dimensional viscous--thermal calculation.  Left: damped density, velocity and temperature amplitudes.  Right: perturbation energy, dissipation and integrated energy-balance residual.}
 \label{fig:visctherm}
\end{figure}

\subsection{Stochastic Gibbs acoustic curvature}

For a one-dimensional prescribed volatility with $\bu=0$, let $\theta_1=\partial_x\alpha$.  Then
\begin{equation}
 \theta_0=-\frac12\theta_1^2,
 \qquad
 \Theta=0.
\end{equation}
For an isentropic ideal gas,
\begin{equation}
 e_1=-c_T^2\theta_1,
 \qquad
 p_1^{\rm M}=-\rho_0c_s^2\theta_1,
\end{equation}
where $p_0=\rho_0R_{\rm g}T_0$ is the base pressure, so the martingale Gibbs residual $\rho_0e_1+p_0\theta_1$ vanishes.  The reversible drift curvature is
\begin{equation}
 e_0=\frac12c_s^2\theta_1^2.
\end{equation}
The numerical evaluation gives residuals below $7\times10^{-18}$.  The same calculation confirms that $[p_1^{\rm M}]=[p]T^{-1/2}$. This is a pressure-state rate, not a stress impulse. A quantity with stress-impulse dimensions appears only as the low-frequency coefficient \cref{eq:Pieff} of an explicitly retained finite-correlation pressure carrier.

\subsection{Low-Mach and pathwise-isochoric limits}

The low-Mach calculation tests acoustic stiffness, while the pathwise-isochoric limit connects the present compressible theory to the companion incompressible solved-volatility formulation of arXiv:2607.25536. In nondimensional variables, with Mach number $Ma$ and $c_s=Ma^{-1}$, the acoustic eigenvalues satisfy
\begin{equation}
 \lambda_\pm=\pm i\frac{k}{Ma}+O(k^2),
\end{equation}
while the slow thermal branch remains diffusive.  The numerical calculation confirms $Ma|\Im\lambda_+|/k\to1$ with error below $3\times10^{-7}$ over the smallest Mach numbers considered.

\begin{figure}[ht]
 \centering
 \includegraphics[width=0.68\textwidth]{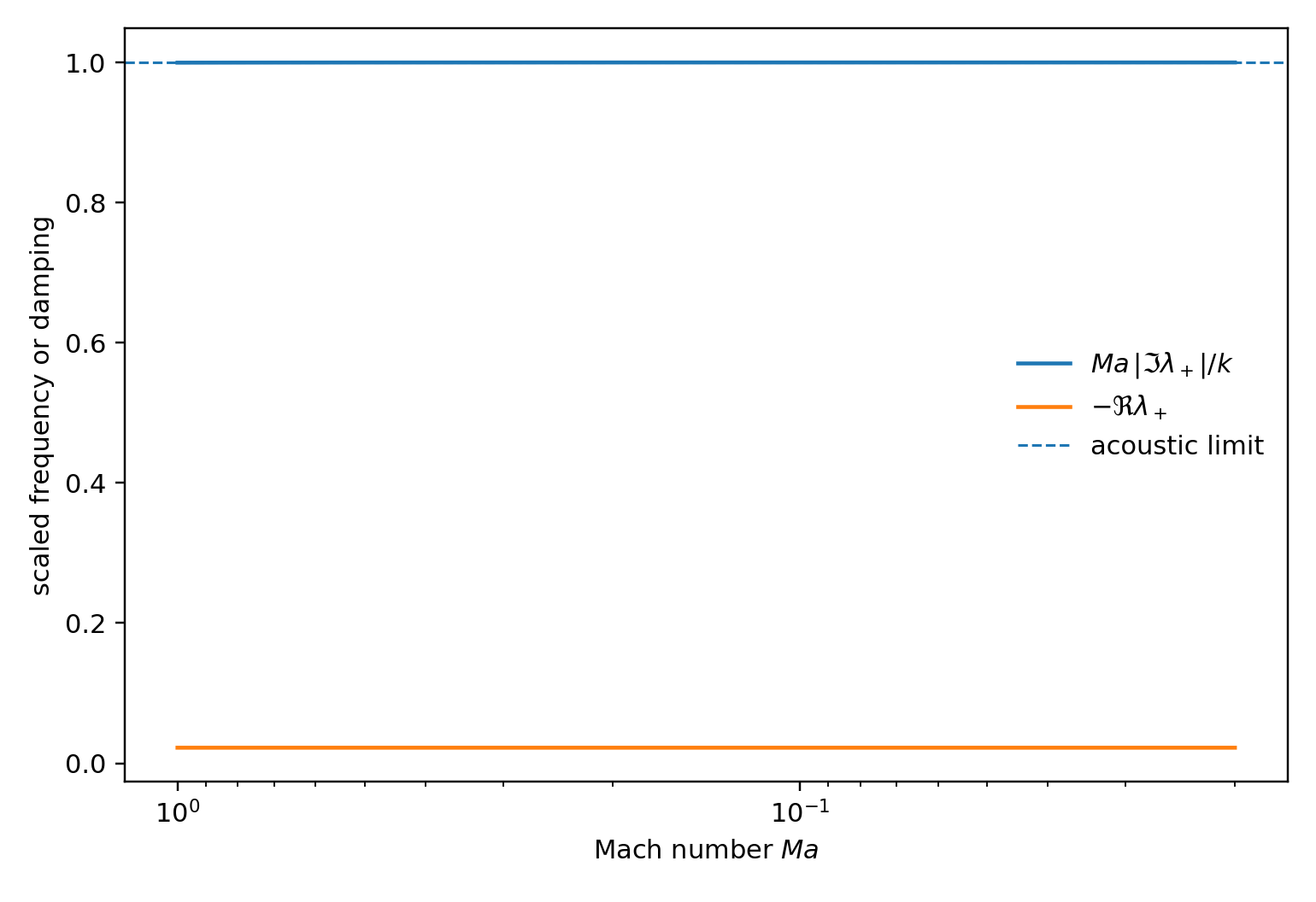}
 \caption{Low-Mach calculation.  The scaled acoustic frequency approaches unity while the viscous--thermal damping remains $O(1)$.}
 \label{fig:lowmach}
\end{figure}

The pathwise-isochoric constant-density limit additionally imposes
\begin{equation}
 \theta_1=0,
 \qquad
 \Theta=0,
 \qquad
 \rho\to\rho_0.
\end{equation}
This is a singular constitutive limit, not a regular identification of the EOS-derived material pressure martingale coefficient with a stress impulse. Thermodynamic pressure loses its equation-of-state role and the mechanical drift and martingale pressures become constraint multipliers. In the augmented carrier description this corresponds to $K_s\to\infty$ and $K_s\tau_p\to\infty$, so that the volumetric fast rate becomes constrained and the limiting martingale pressure $p_1^{inc}$ is determined by the momentum constraint. Provided the reservoir exchange converges in the combined renormalised energy balance, the H--R stress and source-consistent momentum equation recover the incompressible formulation of arXiv:2607.25536 \citep{Tsai2026PaperI}. The conditional compressible entropy result does not create a new white-noise reservoir energy in that limit.

\section{Discussion and limitations}
\label{sec:discussion}

The resulting formulation is thermodynamically more restrictive than a formal addition of density and temperature equations.  Three distinctions are essential.

First, the current-volume rate $\Theta$ differs from the drift of $\log J$.  Bulk pressure work and bulk viscosity must use the actual volume ratio increment.  Second, the source--transport covariance is part of every drift balance and of the associated conservative boundary flux; the displayed martingale terms identify the stochastic transport and impulse channels explicitly.  Third, a white-noise displacement field is not an ordinary finite-energy velocity.  Random--random stress power must be handled at finite correlation time before the limit.

The entropy analysis supports the H--R partition and excludes the raw Hencky power as a pointwise dissipation law.  This retains the It\^o--Hencky stress in momentum while placing the sign-indefinite rate mismatch in a reversible reservoir.  An alternative work-conjugate formulation based directly on $\dev\bD^\star$ is also entropy-admissible but does not preserve the same incompressible constitutive limit.

The pressure typing correction is equally consequential. The equation of state determines the fixed-position pressure time derivative and the material pressure It\^o coefficients, but those rates do not constitute a stress impulse. The finite-correlation carrier \cref{eq:pressurecarrier} shows why a local algebraic pressure-impulse rule is incomplete: pressure--volume work, carrier energy and relaxation entropy form one coupled balance. The regular finite-Mach white-noise baseline therefore contains no intrinsic bulk pressure impulse. A non-zero fast pressure is retained as a coloured carrier, while the incompressible martingale pressure arises only through a singular constraint limit.

The same discipline applies to thermal fluctuations. The present paper does not define a stochastic Fourier law by analogy. A future thermal extension must begin from an augmented conservation law for the thermal carrier and derive, rather than assume, the heat-flux impulse, its source--transport covariation, carrier energy, equilibrium counterterm, entropy and boundary fluctuation law. Only after that derivation may a white-noise heat-flux martingale be introduced without double-counting energy or entropy.

No claim is made for nonlinear well-posedness, entropy solutions with shocks, multichannel rough-path uniqueness, a universal turbulence spectrum, wall logarithmic scaling or computational validation.  The canonical calculations test necessary consistency relations; they do not constitute evidence of developed turbulence.

\section{Conclusions}

A compressible solved-volatility theory can be organised as a mutually consistent chain:
\begin{equation*}
\begin{gathered}
 \text{It\^o--Hencky kinematics}
 \to\text{source-consistent mass/momentum/total energy}\\
 \to\text{constitutive and boundary data}
 \to\text{resolved kinetic energy}\\
 \to\text{internal energy, temperature and EOS pressure}
 \to\text{finite-correlation reservoir}\\
 \to\text{pressure carrier (when retained)}
 \to\text{Gibbs identity and entropy inequality}.
\end{gathered}
\end{equation*}
The ideal-gas specialization recovers the full classical compressible Navier--Stokes--Fourier mass, momentum, total-energy, temperature and pressure equations in the zero-volatility constitutive limit. The finite-Mach conservation subsystem has a mixed and partially dissipative hyperbolic--parabolic symbol. The one-channel drift symbol is directionally rank deficient and must be interpreted together with the martingale compatibility block. Elliptic pressure operators arise only after a specific algebraic closure is imposed or in the singular low-Mach constraint limit. The theory also gives the formal pathwise-isochoric limit associated with the companion incompressible formulation through the singular pressure-carrier and reservoir limits, while making explicit the energy and boundary terms that cannot be assigned directly in white noise.  The H--R formulation gives non-negative predictable entropy production under stated constitutive and detailed-balance assumptions.  For the one-channel formulation considered here, the drift and martingale boundary fluxes stated after the constitutive laws support closed, open, active and passive wall formulations.  The descriptor-symbol, acoustic, viscous--thermal, stochastic Gibbs and low-Mach calculations reproduce the corresponding analytical limits within the reported numerical tolerances.  Open problems include calibration of the pressure-carrier relaxation time and its coupling to the full solved-volatility reservoir, coercivity analysis of the singular low-Mach Schur complement, a conservation-law derivation of any stochastic Fourier extension, nonlinear canonical computations, shock/weak-solution analysis and eventual compressible-flow validation.

\appendix

\section{Boundary-flux derivation}

Integrating \cref{eq:Cs} over a fixed domain gives the generic scalar drift boundary term
\begin{equation}
 \int_{\partial\Omega}\left(q\bu_c+\balpha R_1-\frac12\bSigmaAlpha\nabla q\right)\cdot\bn\,\dd S\,\dd t
\end{equation}
and the martingale boundary term
\begin{equation}
 \int_{\partial\Omega}q\balpha\cdot\bn\,\dd S\,\dd W_t.
\end{equation}
If the source is written as a divergence plus a local supply, its divergence is moved to the left and yields \cref{eq:scalarflux}. Applying the same step to mass, momentum and total energy gives \cref{eq:massflux,eq:momflux,eq:Eflux}.  This derivation explains why prescribing only $\bT_r\bn$ is insufficient on an open stochastic boundary.

\section{Acoustic polynomial and energy symmetriser}

The determinant of \cref{eq:acousticmatrix} is
\begin{equation}
 \det\begin{pmatrix}
 \lambda&ik&0\\
 ikc_T^2&\lambda+\nu_Lk^2&ikc_T^2\\
 0&ik(\gamma-1)&\lambda+\chi k^2
 \end{pmatrix},
\end{equation}
which expands to \cref{eq:acousticcubic}.  The hyperbolic matrix is symmetrised by
\begin{equation}
 \mathsf H=\rho_0\operatorname{diag}\left(c_T^2,1,\frac{c_T^2}{\gamma-1}\right),
\end{equation}
leading directly to \cref{eq:E2,eq:E2diss}.

\section{Numerical verification}

The source archive contains the manuscript, figures, verification scripts and machine-readable numerical results; the file inventory is given in \path{README_SOURCE_v2.0.txt}. The symbolic dispersion residual is zero. The largest observed discrepancies are $7.99\times10^{-8}$ for the acoustic low-$k$ asymptotics, $2.28\times10^{-8}$ for the thermal branch, $1.29\times10^{-5}$ for the integrated energy balance, $2.79\times10^{-7}$ for the low-Mach scaled frequency and $6.94\times10^{-18}$ for the stochastic acoustic-curvature identity. The pressure-carrier calculation confirms $K_s=\gamma p$ for an ideal gas, the exact coupled energy balance \cref{eq:pressurepairenergy}, non-negative relaxation entropy, the low-frequency coefficient \cref{eq:Pieff}, the $|\omega|^{-1}$ high-frequency response and the fast-relaxation bulk-viscous reduction. The maximum relative residual of the sampled carrier-energy identity is below $8\times10^{-16}$.

\section*{Acknowledgements}
The author used generative artificial-intelligence tools for language editing, document organisation, symbolic checks and code assistance.  The author determined the scientific scope, reviewed the derivations and accepts responsibility for the manuscript.

\section*{Funding}
This research received no specific grant from any funding agency in the public, commercial or not-for-profit sectors.

\section*{Declaration of interests}
The author reports no conflict of interest.

\section*{Data and code availability}
The arXiv source archive contains the manuscript, figures and verification code for the canonical limits, ideal-gas specialization and principal-symbol calculations. No compressible CFD dataset is reported because the present work is theoretical.

\end{document}